Article

# The spatially variable effects of mangroves on flood depths and losses from storm surges in Florida

## Graphical abstract

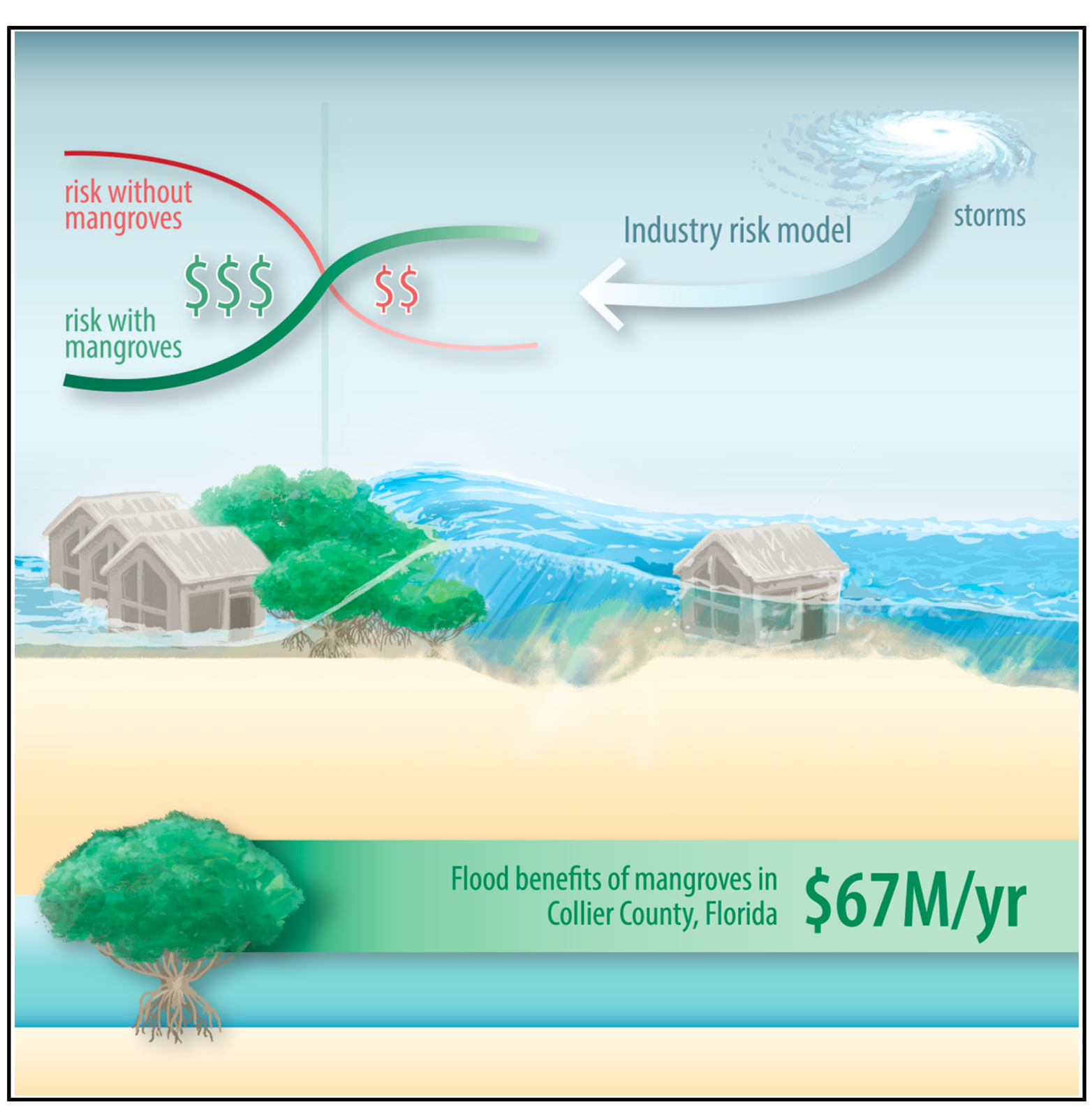

## Highlights

- Mangroves reduce annual storm surge losses in Collier County, Florida, by US$67 million
- 73% of annual mangrove benefits in Collier are for <50-year return period surge events
- Mangroves reduced surge losses in H. Irma by $725 million (14%) and H. Ian by $4 billion (30%)
- Mangroves always reduce landward property losses; effects are mixed for seaward properties

## Authors

Siddharth Narayan,
Christopher J. Thomas, Kechi Nzerem,
Joss Matthewman, Christine Shepard,
Laura Geselbracht, Michael W. Beck

## Correspondence

narayans19@ecu.edu (S.N.),
mwbeck@ucsc.edu (M.W.B.)

## In brief

This study describes the variability in mangrove effects on storm surge losses to properties at spatial scales of kilometers, across multiple storms. Mangroves avoid annual storm surge losses by US$67 million in Collier County, with nearly 75% of these savings occurring for storms with return periods under 50 years. During a storm surge, mangroves always reduce losses for landward properties, whereas for properties inside and seaward of mangrove forests, the effect of mangroves, though net positive, is mixed and can sometimes increase damages.

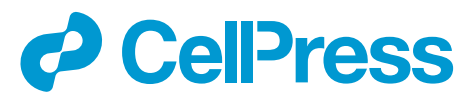

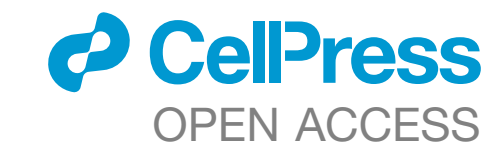

Article

# The spatially variable effects of mangroves on flood depths and losses from storm surges in Florida

Siddharth Narayan,[1,2,6,8,*] Christopher J. Thomas,[3,6] Kechi Nzerem,[3] Joss Matthewman,[3,4] Christine Shepard,[5] Laura Geselbracht,[5] and Michael W. Beck[2,7,*]

[1]Department of Coastal Studies, Integrated Coastal Programs, East Carolina University, Greenville, NC, USA
[2]Center for Coastal Climate Resilience, University of California, Santa Cruz, Santa Cruz, CA, USA
[3]Moody's RMS, London, UK
[4]Reask, London, UK
[5]The Nature Conservancy, Arlington, VA, USA
[6]These authors contributed equally
[7]Senior author
[8]Lead contact
*Correspondence: narayans19@ecu.edu (S.N.), mwbeck@ucsc.edu (M.W.B.)

**SCIENCE FOR SOCIETY** Storm surges from tropical cyclones and hurricanes cause billions of dollars in coastal property damages every year. However, natural ecosystems such as mangrove forests can, by their presence on these coastlines, modify storm surges and affect property damages. Florida in the US is a prime example of a state with an extensive coastline, expensive coastal properties that are exposed every year to hurricanes, and bountiful coastal mangrove forests. Despite growing knowledge of the role that mangroves play in reducing storm surge damages, little is understood of how the magnitude of mangrove effects varies spatially, i.e., where and to what extent properties may benefit from mangrove presence during a storm surge. In this study, we use high-resolution storm surge and property loss models, along with data on mangrove extents in southwest Florida, to quantify the effect of mangroves on property damages annually and during recent Hurricanes, Irma (2017) and Ian (2022). We find that mangrove presence reduces property damages substantially—$67 million annually in Collier County, by 14% during Irma and by 30% during Ian. We also show a strong pattern in the spatial variability of mangrove effects—landward properties always benefit from having mangroves as a natural defense; in situations where properties are inside mangrove forests, they continue to benefit but can, in some situations, especially when they are seaward of these forests, actually face higher damages due to mangroves. These findings underline the importance of understanding the effect of mangrove presence on storm surge damages on coastlines globally, especially when considering their role as natural coastal defenses.

## SUMMARY

Mangroves modify storm surges with impacts on property damages, but these effects vary spatially and by storm intensity. We use high-resolution, kilometer-scale flood and loss models to examine variability in mangrove effects on surge losses to properties, spatially and by storm intensity, in Florida. We estimate that mangroves reduce property surge losses by $67 million annually in Collier County in southwestern Florida. More than 50% of this reduction occurs for storms with return periods under 30 years. Mangroves in Florida reduced storm surge damages by 14% during Hurricane Irma (2017) and 30% during Hurricane Ian (2022). We show that mangrove presence always reduces flood losses for properties landward of mangroves. Inside mangroves, their effect is a net reduction in damages, though in some locations, especially for more seaward properties, mangroves can increase storm damages. These findings underline the importance of mangrove presence in determining property damages during hurricanes.

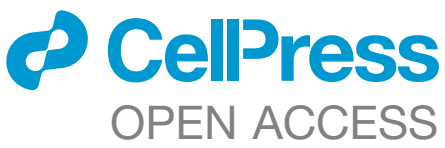

## INTRODUCTION

Mangrove forests have been shown to reduce flood depths during tropical cyclones, thereby reducing flood damages to people and property. This intertidal vegetation acts as a barrier to oncoming storm waves and surges.[1–3] Previous work using statistical models and coarse resolution (>1 km) process-based models has shown that mangroves help reduce economic losses to properties from storm events.[4–6] These benefits are particularly high in places where mangroves are present together with high socioeconomic value in the coastal floodplain and exposure to storm surges. For example, globally, Florida in the USA receives some of the greatest flood damage reduction benefits from mangroves,[4] and in the Philippines, mangrove forests are estimated to reduce around US$1 billion in property damage annually.[7]

The coastline of Florida is a prime example of the combination of storm surge hazards and high economic exposure to these hazards. In 2022, Hurricane Ian made landfall multiple times in southwest Florida on September 28 as a category 4 storm, first in Dry Tortugas National Park in the Florida Keys, in Lee County, and finally, just south of Punta Gorda in Charlotte County, Florida (Figure 1). The storm surge from Ian, the first category 4 hurricane to impact southwestern Florida since Charley in 2004, impacted barrier islands and coastlines from Fort Myers Beach to Naples Bay.[8] Five years earlier, in 2017, Hurricane Irma devastated nations in the Caribbean and Florida, with damage costs exceeding US$200 billion.[9] Hurricane Irma made landfall twice in Florida on September 10, 2017, first in the Florida Keys as a category 4 storm and later in Collier County as a category 3 storm (Figure 2). Some of the highest storm surges of 8 to 10 ft were observed along the mangrove-dominated coastline of the Everglades National Park.[10]

While general awareness of the overall benefits of mangroves for coastal protection during storms has been growing, knowledge of the net economic value of mangroves for reducing storm surge damages to properties from multiple storms remains incomplete, particularly our understanding of the variability in these effects at kilometer scales, both spatially, and in terms of how these benefits vary by storm severity. Measurements of the economic effects of mangrove forests on property damages during storms are rare and have mostly been statistical model studies at scales of 10–100s of kilometers or more, based on hurricane wind speeds and broadly defined storm impact areas.[5,6,11] Coastal and freshwater wetlands have been estimated in one study to have reduced county-wide damages from past storms in Florida anywhere between US$5,000 to US $1,617,000 per year.[6] Using a statistical model of the effects of wetlands on county-level property damages from storms between 1996 and 2016,[6] it is estimated that a loss of 500 km$^2$ of wetlands in Florida—mostly mangroves—would have increased losses during Hurricane Irma by US$430 million and that, across the US, wetland protective effects were greater for lower category storms. Another global statistical model estimated annual avoided loss values for coastal and freshwater wetlands of over US$100 million over 100-km by 100-km areas in most of Florida.[11] These studies improve our understanding of the effects that wetlands can have during hurricanes but stop short of describing how these effects vary by event frequency or spatially, particularly at kilometer scales, partly due to a lack of data on variability in observed damages at these finer scales. Spatial variations in wetland values include, for example, the possibility of higher damages due to the possibility of tidal vegetation increasing flood depths from storm surge flows, as has been shown using 2D hydrodynamic models for salt marshes in the US Northeast, Pacific coast, and Texas.[12–14] Combining state-of-the-art physics-based models with property damage estimates, our study advances current understanding of mangrove presence on property losses from storm surges at kilometer scales by (1) improving quantitative estimates of the net coastal protection benefits of mangroves on flood depths and property losses during storm surges; (2) describing how mangrove effects vary over multiple storm events of varying severity; and (3) describing spatial variations in mangrove effects for two recent events, Hurricane Ian and Hurricane Irma, including where mangroves may have negative effects on property damages.

To explore variability in mangrove effects spatially and by event severity, we focus our analysis on the state of Florida and use a catastrophe risk model widely used in the insurance industry to simulate storm surge flooding and damages from a range of storms. Here, we use a numerical catastrophe risk model specifically designed to assess property-level storm surge losses for two mangrove scenarios: a present-day mangrove wetland scenario and a counterfactual scenario that corresponds to mangroves being replaced with developed open space or an open-water seabed (see methods). For each storm event, we integrate a physics-based model that simulates the generation of storm surge at the coast and the subsequent overland flooding, with an economic loss model at high resolution (~100 m in the coastal floodplain; see methods). For each event and each mangrove scenario, we estimate the resultant property damages and then compare damages across the two scenarios. First, we quantify the effects of mangroves on average annual losses (AALs) from storm surge in Collier County in southwest Florida, by running simulations for a set of synthetic storms chosen to represent 100,000 years of tropical cyclone activity in the area. We also analyze the effect of mangrove presence on storm surge damages for two recent historical tropical cyclones—Hurricanes Irma (2017) and Ian (2022) across Florida.

## RESULTS

### Mangroves reduce surge flood depths and damages

The presence of mangroves in southern Florida reduced flood extents in several locations during Hurricane Ian, Hurricane Irma, and across multiple storms in Collier County. During Ian, the presence of mangroves reduced flood depths by over a meter in several coastal regions of Collier, Lee, and Charlotte Counties (Figure 1). During Hurricane Irma, mangroves reduced storm surge flood depths across the southwestern Florida coastline and in parts of Miami-Dade County in the east (Figure 2).

Our analyses show that by modifying storm surge depths, mangrove forests in southern Florida significantly reduce property damage from storm surges across the entire floodplain. In our model, the economic damages in Florida from storm surges alone were ~US$13 billion during Hurricane Ian and ~US$5 billion during Hurricane Irma (2018 US$). Mangroves in southern Florida

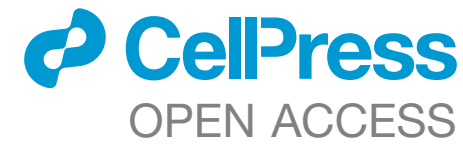

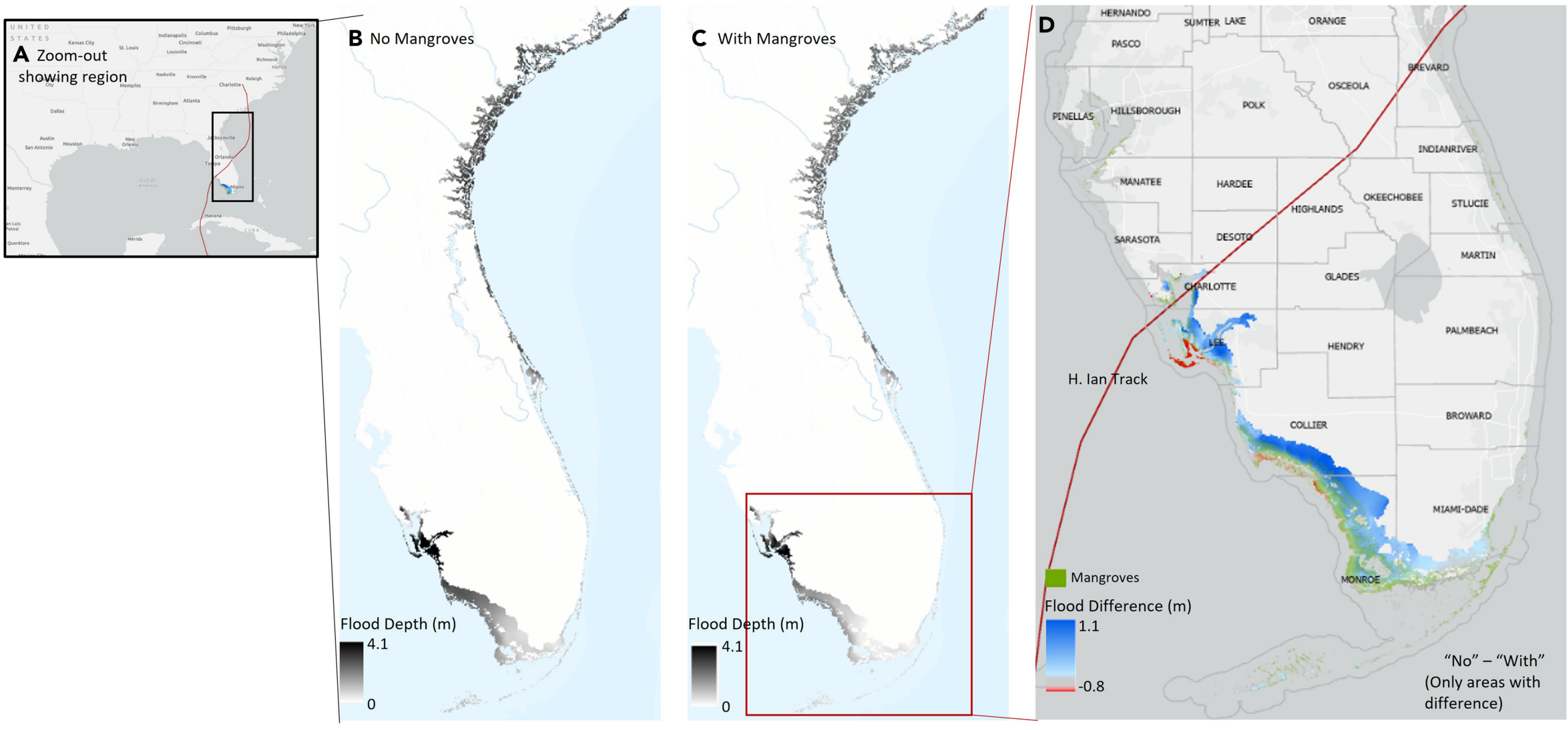

**Figure 1. Flood depths during Hurricane Ian in meters with and without mangroves**
(A) Zoom-out showing region of interest; (B) No-Mangroves scenario; (C) With-Mangroves scenario; (D) Detail inset of Difference (No Mangroves minus With Mangroves) for southern Florida and Hurricane Ian track in red. Figure created using ArcGIS.

reduced damages by ∼US$725 million during Irma, a 14% reduction in damages. During Ian, the complete absence of present-day mangroves would have increased damages by an estimated US $4.1 billion, ∼30% of total storm surge damages, though this effect was felt across a smaller spatial footprint. The approximately 250 km$^2$ of mangroves in Collier County in our model reduce annual losses from storm surge by US$67 million compared with a scenario where mangroves are absent, a net benefit of ∼US$270,000 per km$^2$ per year of mangrove forest.

## Mangrove benefits are greater for more frequent storms

Across multiple events in Collier County, our study shows that mangrove effects vary by storm event severity and are cumulatively more significant for less damaging storm events that are

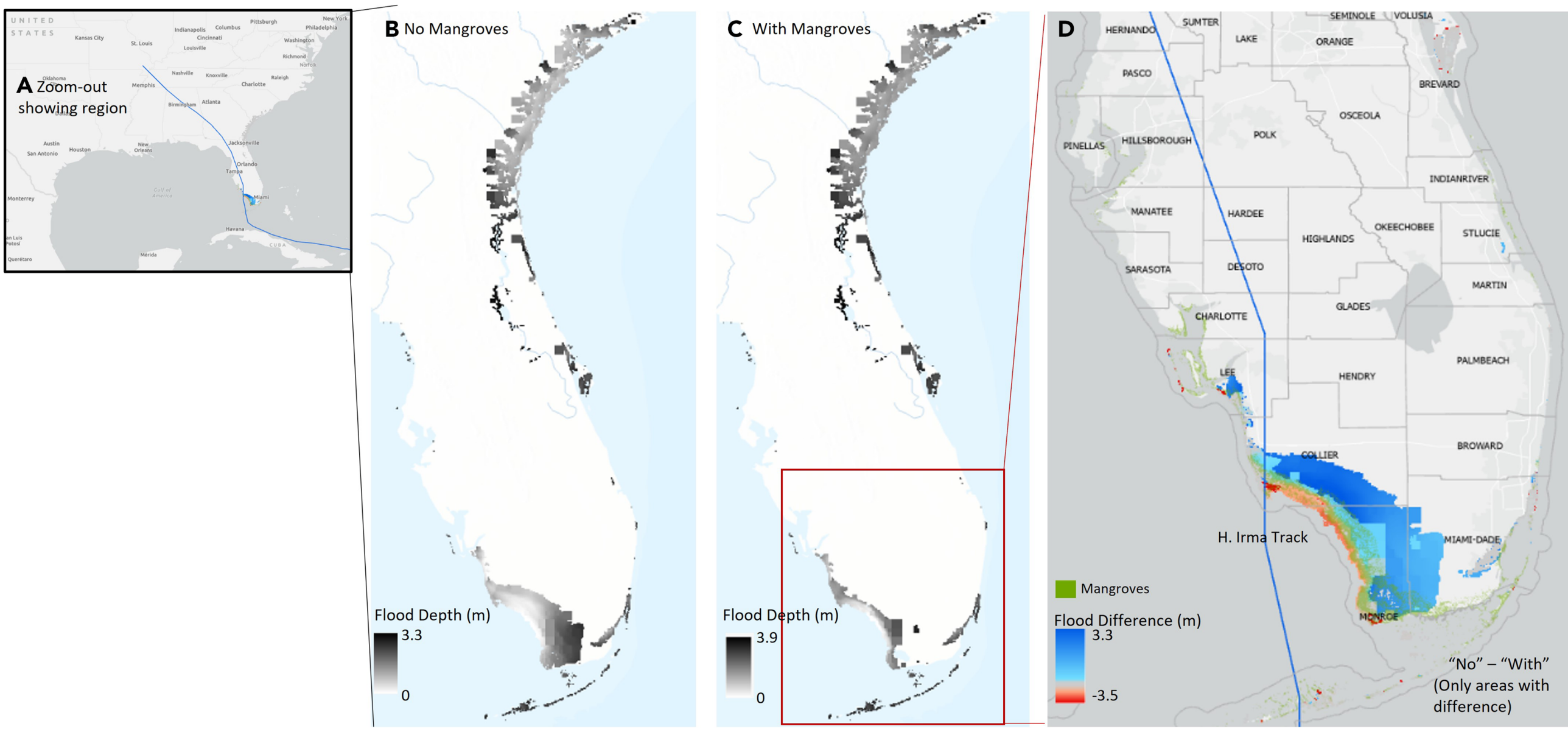

**Figure 2. Flood depths during Hurricane Irma in meters with and without mangroves**
(A) Zoom-out showing region of interest; (B) No-Mangroves scenario; (C) With-Mangroves scenario; (D) Detail inset of Difference (No Mangroves minus With Mangroves) for southern Florida and Hurricane Irma track in blue. Figure created using ArcGIS.

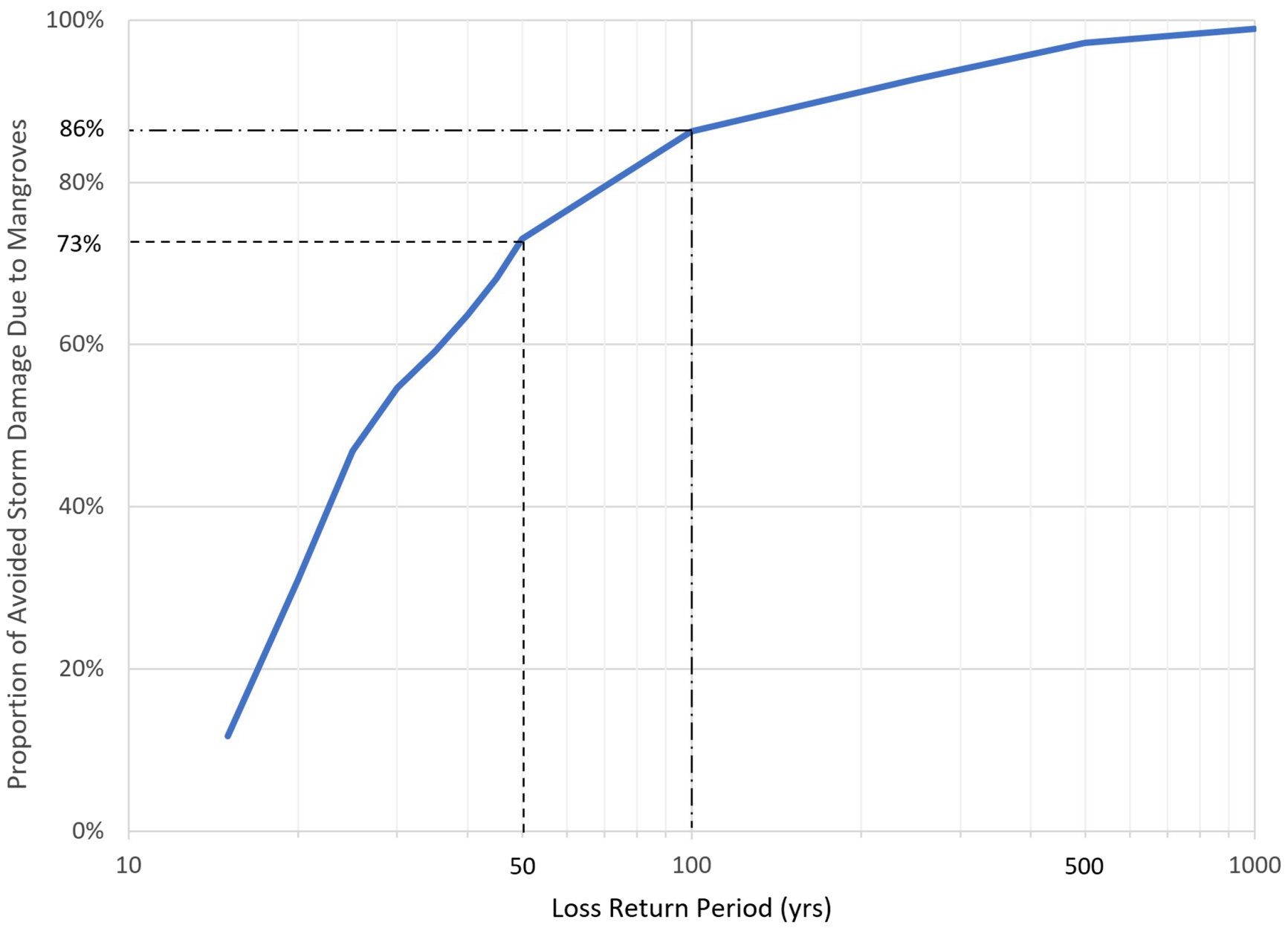

**Figure 3. Cumulative proportion of AAL reduction benefits from mangroves by storm return period**

Vertical axis indicates the total annual benefits (equal to US$67M) coming from storms with loss-based return periods (RPs) up to and including the value given on the *x* axis (log scale). For example, a *y* axis value of 73% corresponds to an *x* axis value of 50 years, meaning that 73% of the US$67M AAL reduction comes from events up to a 50-year loss-based RP. Dashed line indicates a 50-year RP event and dot-dash line indicates a 100-year RP. Values for RPs below 15 years are not plotted due to high model uncertainty at very low RP values; the upper *x* axis is clipped at 1,000 years for clarity (see methods and Figure S1).

typically more frequent (Figure 3). Over half (55%) of the total mangrove-induced reduction in AALs in Collier County occurred for storm surge events with loss return periods (RPs) below 30 years, i.e., events with an annual loss frequency above 3.3%. Almost three-quarters (73%) of the cumulative reduction in AALs by mangroves are within the 1-in-50-year loss RP, whereas only 14% of these benefits are for events with loss RPs greater than 100 years, i.e., an annual loss frequency below 1%.

## Mangrove effects vary between landward and seaward properties

Mangrove effects during Ian were concentrated in Charlotte, Lee, and Collier counties, the three counties closest to where Ian made landfall on the mainland in Charlotte County (Figure 4B). During Irma, mangrove effects were more widely distributed across southern Florida (Figure S2).

Mangrove effects also varied spatially within the study domain, annually for Collier County (Figure 5), for Hurricane Ian (Figure 5), and for Hurricane Irma (Figure S2). The spatial variations in mangrove effects exhibited a zonation depending on whether the properties were farther inland from the landward mangrove edge or inside the mangrove forests and seaward of this edge. Mangroves reduced storm surge property losses for 100% of hexagons inland of the landward mangrove edge for Collier County AALs, for Hurricane Ian (Figure 5) and Hurricane Irma (Figure S3).

For the AALs in Collier County, mangrove effects are significant, i.e., defined as being greater than 1% of the absolute maximum effect of mangrove presence, for properties located up to 11 km from the mangrove edge (Figure 5). During Ian, mangrove effects were significant as far as 13 km landward of the forest edge (Figure 5). During Irma, mangrove effects were significant within 14 km of the forest edge for landward properties, and up to 5 km from the edge for properties between mangrove forests (Figure S2).

The net effects of mangroves for the AALs and Hurricane Ian were positive, and greatest in magnitude, for all properties landward of the mangroves (Figure 6). These effects were net positive but much smaller for properties inside the mangroves. Like the flood depths, the effects of mangroves on economic losses for Hurricanes Ian and Irma were mixed for properties inside the mangroves. For example, on the outer islands where Ian made landfall, some properties received benefits from the mangroves, while adjacent properties saw an increase in damages due to mangrove presence (Figure 5). This is likely due to the mangrove vegetative barrier causing an increase in water levels in some areas within and seaward of the forests and these regions coinciding with areas that have properties. For Hurricane Irma, despite this effect, properties inside the mangroves saw a greater reduction in total storm damage than properties landward of the mangrove edge, possibly due to the overall larger number of exposed properties within the mangroves.

## DISCUSSION

Our study estimates, at high spatial resolution, the net benefits of mangroves during multiple hurricane events as well as the variability in these effects on storm surge damages to properties. We find that a majority of the protective benefits from mangroves accrue for smaller, more frequent storm surge events; that the net effect of mangroves is highly beneficial despite some negative effects at specific locations; and that, in addition to cyclone characteristics, the benefits and negative effects of mangroves depend on the location of properties relative to the mangrove forests and the coastline. Together, our results indicate that mangrove forests have a high economic value as natural defenses, especially for properties landward of them during smaller storm events. The net economic benefits of mangroves strengthen the argument for considering them as national natural infrastructure for coastal risk reduction.[15]

Our results describe the finest-scale estimate to date of the effects of mangroves on storm surge property damages from multiple storm events, indicating that mangroves reduced storm

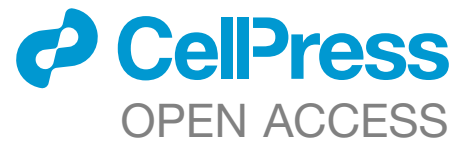

**Figure 4. Spatial variability in storm surge damage differences with and without mangroves in Collier County and for Hurricane Ian**

(A) (Top) Difference in storm surge damages between the No-Mangroves and With-Mangroves scenarios for AALs in Collier County. Mangroves are shown in green and landward mangrove edge shown as a dark green line. Areas behind this edge are in solid hexagons, areas inside and seaward of this edge are in dashed hexagons. Red-shaded hexagons see increase in damages due to mangroves; blue-shaded hexagons see reduction in damages due to mangroves.

(B) (Bottom) Difference in storm surge damages between No- and With-Mangroves scenarios for Hurricane Ian. Figure created using ArcGIS.

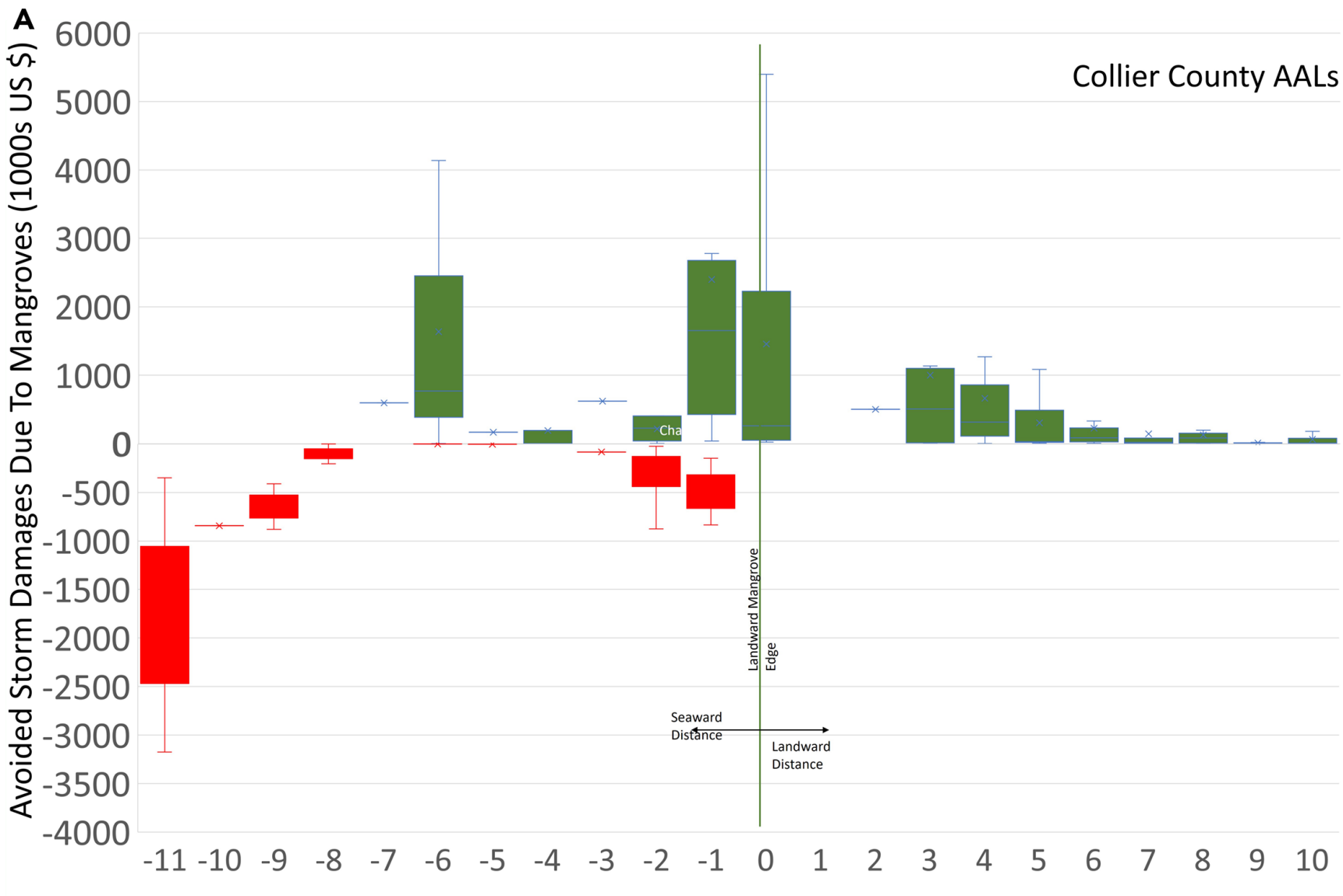

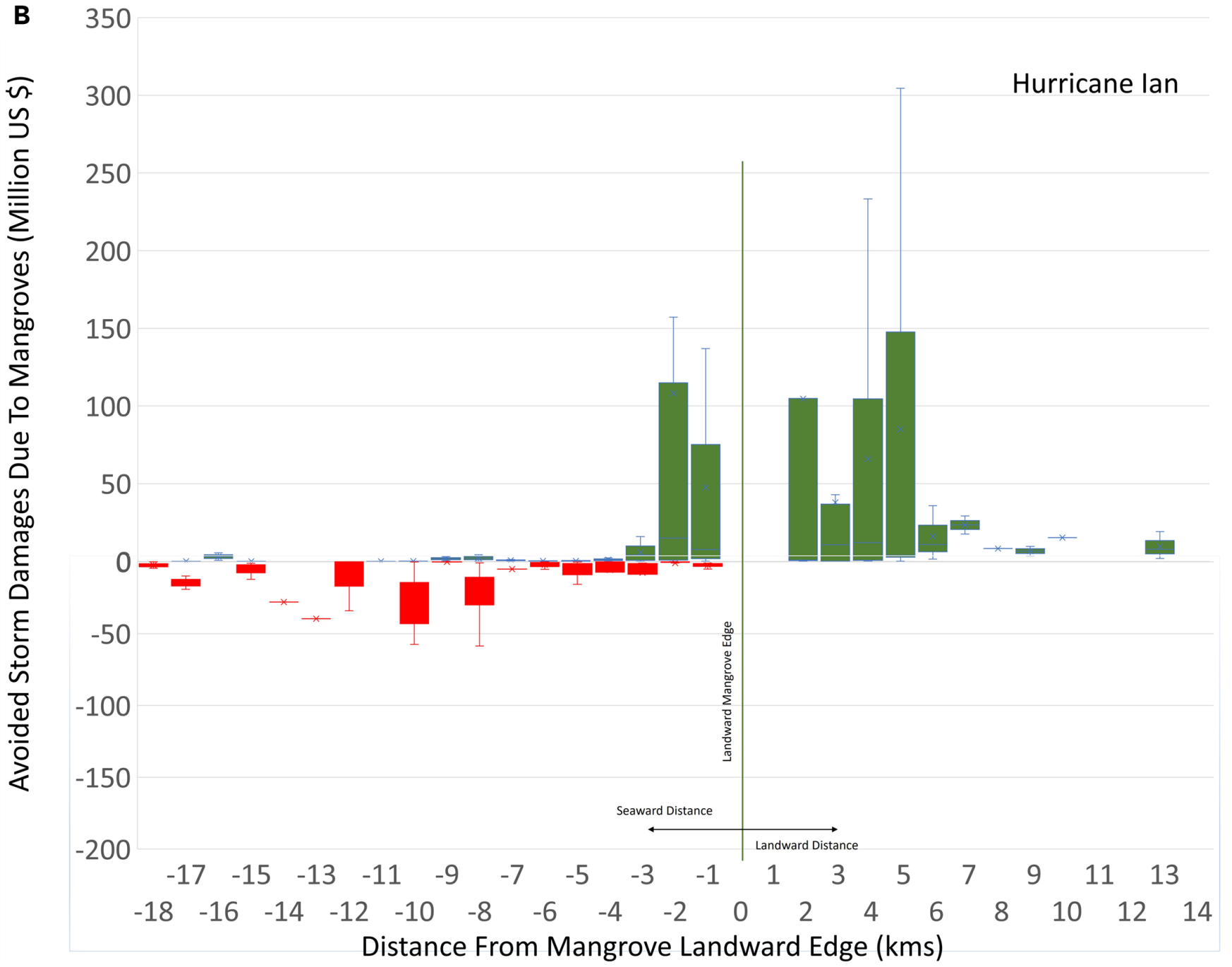

**Figure 5. Storm surge damage differences due to mangroves by distance**

Box and whisker plot showing difference in storm surge damages between With-Mangroves and No-Mangroves scenarios for 1-km distance bins from the landward edge of the mangrove forests for (A) (Top) Collier County AALs, and (B) (Bottom) Hurricane Ian for southern Florida. Green bars show reduction in risk due to mangroves and red bars show increase in risk due to mangroves. Green line indicates landward mangrove edge. See Figure S3 for Hurricane Irma map and plot.

surge damages between 14% to 30% during Hurricanes Irma and Ian. In Collier County, by analyzing 100 storm events we estimate that mangroves reduce average losses from storm surges by US$67 million annually, which translates to a benefit of ∼US $270,000 per km$^2$ of mangroves per year, an effect that is highly spatially variable (Figure 5). This value is an order of magnitude higher than the US$38,000 per km$^2$ per year value estimated by Sun and Carson[6] for Collier County, who considered effects of coastal and freshwater wetlands for 15 storm events between 1998 and 2010 of which only four caused observed damage to the county. One possible reason for our per-km$^2$ estimates being higher is the larger storm dataset in our study, which analyzes 100 synthetic storm events, all of which cause flood damages in Collier County. Another possible reason is the larger area of coastal and freshwater wetlands considered in Sun and Carson[6]—between 1,400 and 2,500 km$^2$, depending on the storm. The protective value of wetlands for wave dissipation has been shown to be non-linear with wetland extent, with the first few kilometers of wetlands providing the largest benefits, with a similar effect likely for storm surge dissipation.[16] On the other hand, our estimates of mangrove value during Hurricane Irma are lower than indicated by Sun and Carson,[6] while our study estimates that all mangroves across southern Florida in total reduced damages by US$725 million, their model suggests that a loss of just 500 km$^2$ of these wetlands could have increased damages by US$430 million, implying that a loss of all mangroves in southern Florida would be several times higher. While we do not calculate the effect of historic losses in mangrove area on damages from Irma, it is likely that mangroves in some areas with higher exposure, such as in Broward County, were a lot more valuable than mangroves in other areas, such as the Everglades, that do not have much property exposure behind them (Figure S2). In general, our findings underline the importance of more detailed studies to improve these estimates and better describe variability in these effects.

Our results suggest a temporal, event-frequency scale to the accumulation of mangrove benefits for risk reduction in Collier County, similar to effects found by previous studies. Over half of the annual loss reduction benefits from mangroves accrue for storm events with RPs of under 30 years, benefits that could be realized within the lifetime of a typical home mortgage in the US. Nearly three-quarters of the total annual benefits of mangroves are received for events with a RP below 50 years. This could be because storm surges from smaller events typically

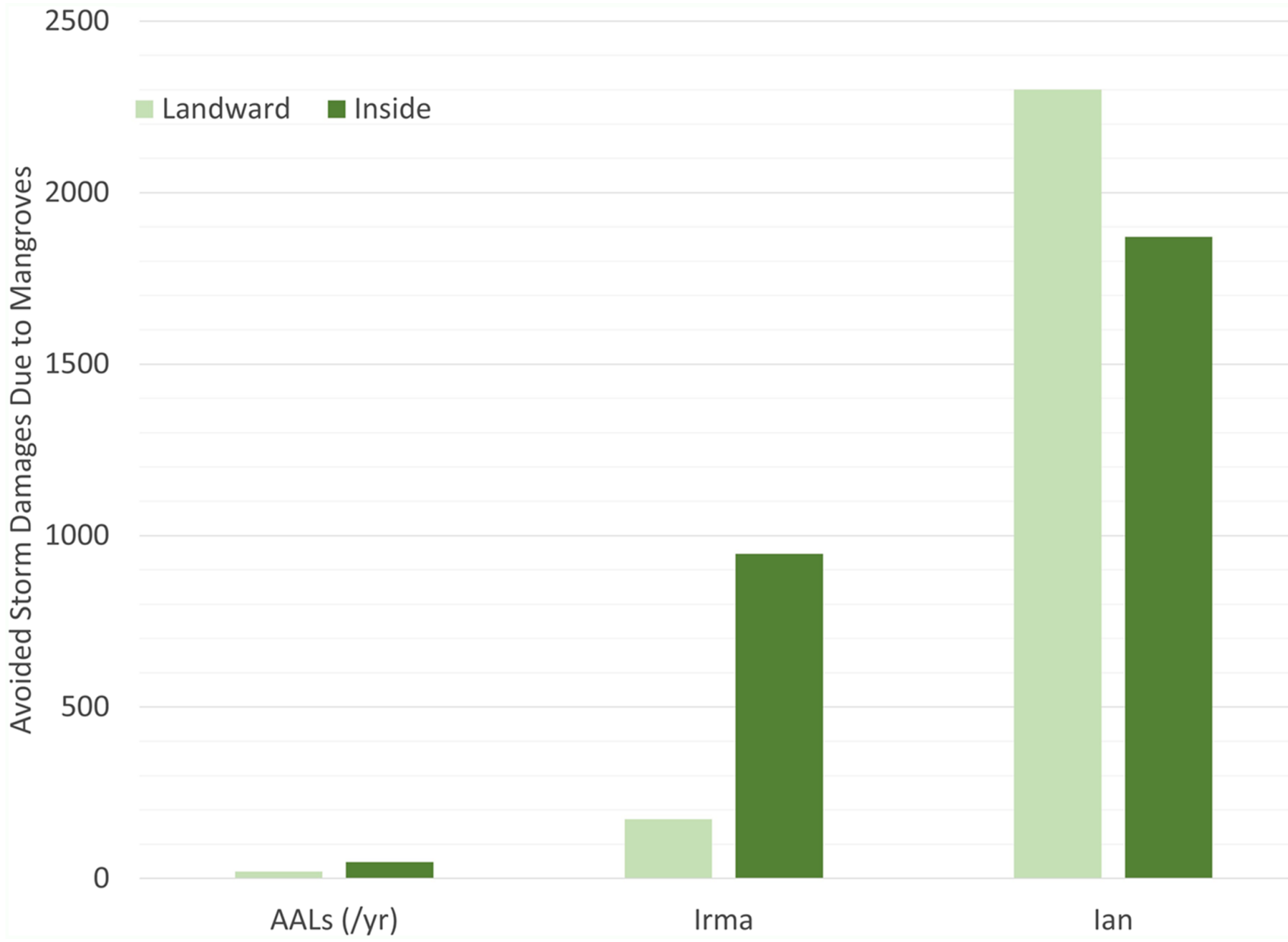

**Figure 6. Net mangrove benefits for the three event cases landward and inside mangroves**

Benefits for AALs (million US$s/yr) in Collier County, and across southern Florida for Hurricanes Irma (million US$s), and Ian (million US$s), in two zones: landward of mangroves (light green); inside and seaward of mangrove areas (dark green).

have lower inundation depths and are therefore more effectively slowed down and redirected by the friction provided by the mangrove vegetation.[17] The contribution of mangroves during smaller events is significant, as we find that within our simulated storm set, 56% of total damage is caused by events with an RP < 100 years (Table S1). This supports the finding by Sun and Carson,[6] whose US-wide model suggests that wetland effects are cumulatively greater for smaller events. The benefits of mangrove forests for smaller, more frequent storm events could help incentivize homeowners, who play an important role in coastal adaptation in the US, to support the consideration of natural ecosystem alternatives within adaptation response portfolios.[18,19]

Our study presents an initial exploration of a less-studied aspect of mangrove effects during storm surges, i.e., the spatial variability of these effects especially for property losses inside and seaward of mangrove forests by describing the 2D process of storm surge propagation through mangroves.[3,17] In general, our model illustrates a zonation of these spatially variable effects, with landward properties having entirely net positive effects with lower spatial variability, whereas the effects are much more spatially variable for properties that are inside and seaward of these forests, a pattern that holds true for multiple events in Collier County (Figure 5). There is noticeable spatial variability in the effects of mangroves on surge damages on properties in the outer islands for AALs in Collier County and similarly for Hurricane Ian (Figure 5) and Irma (Figure S2). This is due to the mixed effect of mangrove patches in a region very close to landfalling hurricanes and their accompanying surges, with some areas experiencing a reduction in flood depths, and others an increase in heights due to the mangroves (Figures 1 and 2). These results support recent studies that indicate similar effects in salt marsh wetlands, a temperate intertidal analog to mangroves.[12–14] As the evidence for mangrove forests as coastal defenses, and consideration of these nature-based alternatives to hard structures, increases[20] more detailed studies are needed to better understand how these natural defenses alter storm surge risks to properties around them, and how this changes by storm severity and duration.

Our flood model performs well in estimating storm surge flood depths, validating our approach and choice of mangrove land-cover coefficient values. Our model is validated against observed the US Geological Survey (USGS) high-water marks during Hurricane Ian and Hurricane Irma and shows a mean error of 0.1 m and a root mean square error of 0.5 m or less in both cases (Figure 7) for flood depths higher than 2.5 m during Irma and 3.5 m during Ian. Our model uses a Manning's friction coefficient approach based on publicly available data on land-cover classes (Figure S4), with a coefficient of 0.1 to simulate the effect of mangrove extent on the flow of water over land. The Manning's n approach has been widely used and validated for describing land-cover effects on storm surge propagation, and our value for mangroves is within the range found in previously published, validated storm surge models.[21,22] Friction coefficient values for mangrove land cover in previous studies have varied between 0.1 to 0.15 in most cases.[23,24] Our value of 0.1 for mangroves represents a conservative choice, i.e., it is more likely to underestimate rather than overestimate the effect of mangroves on storm surge flows, as indicated by recent studies, including controlled experiments by the US Army Corps of Engineers.[25,26] Thus, the results presented in this paper can be interpreted as a conservative, i.e., lower-bound estimate of the effect of mangroves on flood depths and losses.

Recent advances, such as the use of dynamic coefficients or 3D hydrodynamic models that consider spatial as well as temporal variations in mangrove characteristics and storm surge flow, can provide better characterization of mangroves for future studies,[27,28] for example, by accounting for changes in drag as water depth increases or decreases during a surge and covers a greater proportion of the tree heights.[29] However, these models are accompanied with challenges pertaining to the proper characterization of coastal vegetation structure,[30] as well as increased runtimes. Our model implicitly includes the effect of waves on total water levels to increase the storm surge water level in grid cells exposed to wave attack by adding 70% of the expected significant wave height to the still water depth at these locations. In our flood model we do not change

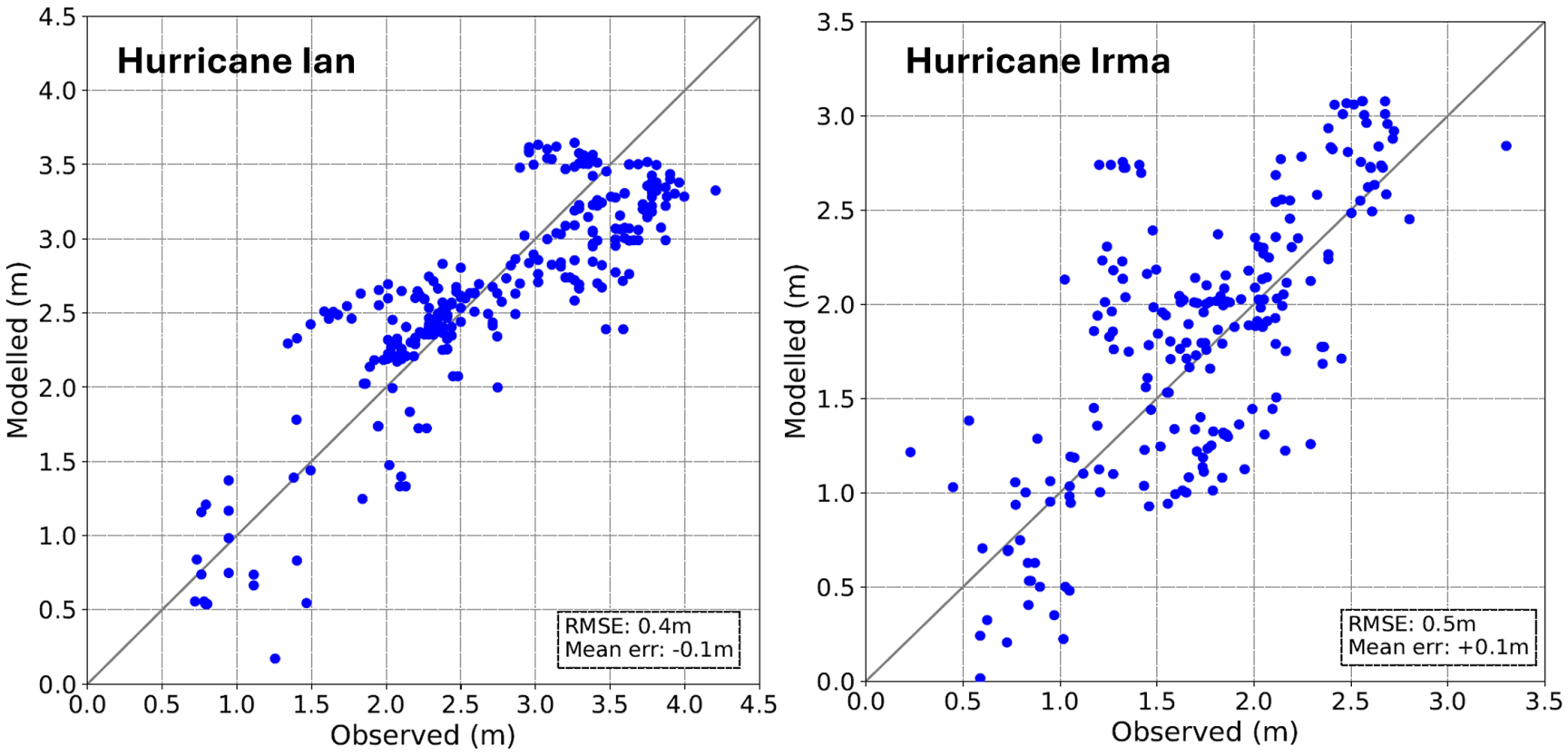

**Figure 7. Observed vs. modeled peak flood elevations in m for Hurricanes Ian and Irma**

(Left) Hurricane Ian and (right) Hurricane Irma. Observed flood elevations are obtained from USGS high-water mark and National Oceanic and Atmospheric Administration (NOAA) tide gauge data.

the locations that are susceptible to wave attack when we change mangrove extents. In further studies, explicit inclusion of the effect of mangroves on wave height contributions to total water levels will provide richer information on the contribution of mangrove forests during a storm, where the inclusion of wave-induced damage effects could potentially increase the contribution of mangroves to avoided losses.[31] Our model uses the best available mangrove extent at the time of the study. Since then, advances in and increased availability of remote sensing data through satellite imagery have resulted in significant improvements in the coverage and resolution of data on mangrove extents.[32] Likewise, having more detailed large-scale spatial data showing the density and health of mangrove forests would also allow us to vary the drag coefficient within mangrove forests, which would be more realistic than assuming a single value.

For Hurricanes Ian and Irma, our storm surge damage estimates fall well within the total estimated damages from flooding across Florida from other reports.[8,10,33] For all hurricane events, including the 100 events in Collier County, our damage model uses the results of the validated flood model to estimate damage from each storm surge to all insurable flooded properties, including time element losses but excluding losses to public infrastructure. In contrast to previous county-scale statistical model studies, our damage model is based on spatially variable flood depths at kilometer scales produced by the flood model (Figures 1 and 2) that are critical for further damage assessments. Storm surge damage assessments, in general, are difficult to estimate and even more difficult to validate. For example, the US National Weather Service (NWS) event database estimates soon after Ian recorded a total damage in Florida of approximately US$2 billion,[34] whereas later reports assess damages at greater than US$110 billion.[8] Hurricane Irma is estimated to have caused over US$50 billion in total damages,[10] much of which was due to rainfall and wind. This estimate includes physical damages to private and municipal buildings and infrastructure, crops and livestock, and time element losses, among others.[35] In comparison, flood-only claims data for damage from Hurricane Irma from the US National Flood Insurance Program (NFIP) show that the total amount of insured flood losses that NFIP paid out for Irma to date is much lower than estimated total damages at just over US$1.1 billion.[36] However, this represents an incomplete picture of total economic losses from flooding during hurricane events.[33] We do not compare our loss estimates to Federal Emergency Management Agency (FEMA) flood insurance claims due to differences in the types of losses counted: our dataset estimates total economic loss to all insurable properties in the region and includes time element losses, while the NFIP claims data only represent claims paid out to NFIP policyholders and thus exclude the value of damage to properties that were uninsurable, uninsured, or underinsured against flood relative to their total value, or for which claims were not filed or paid out.[37] Future improvements to damage models could include, for example, detailed considerations of different structure types and the damages associated with flooding for these typologies.[38]

Our study highlights the importance of up-to-date information on mangrove extents and changes to mangrove extents for the assessment of storm surge losses and similar coastal risks. For our No-Mangroves scenario, we use a lower friction coefficient that corresponds to developed open space such as a golf course, or open water based on previous storm surge assessments in Florida.[21,26] While we do not expect all mangroves to be converted to these land-cover types, our analysis uses this scenario as the best way to quantify the effect of the absence of these mangroves. The development of more realistic counterfactuals, such as possible development over the mangroves, is beyond the scope of this study but comes with its own challenges, such as accounting for the increased exposure to storm surge damages due to coastal development.[12] We also do not consider in this study detailed descriptions of damages to the mangroves themselves from these storm surge events. Previous work has shown that proper characterization of surge risk, land cover, and building types is important for appropriate risk assessment.[39] Mangroves in southwest Florida largely withstood wind forces and did not topple during Hurricanes Irma and Ian, which implies that the overall benefits we estimate likely did not decline during these events. However, mangroves did die off after the hurricane particularly in areas with poorly drained soils, for example, in places where construction restricted tidal flows,[40] which suggests that future flood risk could increase from the loss of these natural defenses, and that proactive and adaptive management of these natural defenses is important.[41]

Recent work shows that while the natural capital benefits of many resources are declining globally, the natural capital benefits of mangroves for flood risk reduction is increasing for many nations around the world with real benefits for national gross domestic product (GDP).[42] Overall, coastal risk is increasing rapidly due to increasing coastal development.[12] The high net benefits of mangroves in these regions are driven by the high value of properties they protect, while the development of some of these properties occurs at the expense of these mangroves.[43] Consequently, the value of mangroves per hectare is increasing largely because of the rapid increase in coastal risk; anything that helps lower this risk has rising value and consequently higher benefit to cost ratios for restoration.[44] We use an insurance industry-standard catastrophe risk modeling approach to measure the annual benefits of mangroves for reducing damages to properties, which can enable opportunities to create incentives for these natural defenses.[45] Insurance risk industry models are used widely by clients from businesses to government agencies for a wide range of considerations including risk assessment and pricing, bonds that offer risk reducing and resilience building measures, and to identify when current or planned measures may offer benefits significant enough for a variety of incentives, such as insurance premium reductions or infrastructure investments. By using risk industry models for quantifying the annual benefits of mangroves, this paper advances the consideration of mangroves as a viable risk reduction strategy within risk insurance practice.

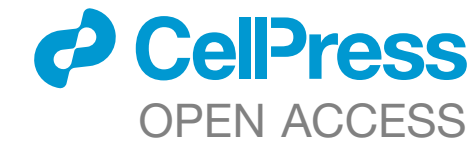

Overall, our study explores at kilometer-scale resolution the effects of mangroves to properties during storm surges across event frequencies, and the spatial variability in these effects. Mangroves appear to be very effective in reducing losses for landward properties and cumulatively more effective for smaller storm surge events. Our study suggests a more limited role for mangroves in reducing damages during much larger events, which may include Hurricane Helene in 2024 when surge heights appear to have exceeded 4.5 m in parts of coastal Florida,[46] though any avoided damage values could still be high, simply due to the extensive damage that such large storm surges can produce.[6] More than US$100 billion was appropriated by the US federal government to recover from coastal impacts after the 2017 hurricane season, while in south Florida alone, mangroves provided over US$4 billion in avoided damages during the 2017 and 2023 hurricane seasons. Few of these recovery funds have so far been used to restore or to repair the flood-reducing natural infrastructure damaged during these storms. This is consistent with past events such as Hurricane Sandy, when less than 3% of recovery funding supported green infrastructure.[47] By assessing the types of storms where mangroves are most beneficial for property loss reduction and spatial variations in these benefits, we can provide evidence to inform public and private investments in the conservation and restoration of this national natural infrastructure for risk management.

## METHODS

We estimate the extent to which mangroves reduce flooding and related property losses from storm surges for two case-studies in southern Florida: (1) annually, using a set of synthetic storm events in Collier County, chosen to represent 100,000 years of tropical cyclone activity in this area; and (2) a reconstruction of the storm surges from Hurricanes Irma (2017) and Ian (2022) across southern Florida. We calculate mangrove benefits by combining multiple datasets and models, using a stepwise approach from the source of the hazard (storm surge), to the receptors of damage (flooded properties), through an intervening pathway (mangroves), and the consequence of this damage (property loss).[48] For each storm event, the effect of the intervening mangroves on storm surge property loss is calculated as the difference between two loss values: (1) With Mangroves; and (2) No Mangroves, i.e., a scenario where mangroves are absent.

### Hurricane parameters

Hurricane tracks and parameters for the multi-event analyses for Collier County are obtained from the large synthetic storm event set contained in the Moody's Risk Management Solutions (RMS) 2018 North Atlantic Hurricane model. This track set was constructed using a statistical tropical cyclone track model covering the North Atlantic Ocean and involves stochastically extrapolating the Hurricane Database (HURDAT) catalog of observed hurricane activity,[49] using the statistical techniques described in Hall and Jewson[50,51] to generate a set of storms representing 100,000 years of hurricane activity in the basin. This synthetic storm catalog spans the full range of what is considered physically realistic while retaining similar statistical characteristics to the observational HURDAT track set (see Hall and Jewson[51,52]). Each storm's track is simulated from its formation to its dissipation, using a semi-parametric model based on historical data.[52] Storm wind fields are constructed using an analytical wind profile derived from Willoughby et al.,[53] with parameters fitted from the extended best track dataset[54] and wind fields from RMS Hurricane Wind (HWind).[55–57] To assess annual losses in Collier County specifically, we select the 3,966 storms in the full synthetic set that impact the county and further distill these into 100 representative storms, which recreate the distribution of surge-related losses in the full storm set, using the methodology described further down in property losses from storm surge flooding.

To simulate the storm surges of Irma and Ian, their wind fields were re-created for this study using event-specific 6-h snapshots generated by RMS HWind, a standardized, observation-based wind field analysis compiled using data from a wide variety of observational sources.[57] A pressure field is constructed following Holland,[58] and these wind and pressure fields are used to drive the flood model which simulates storm surge flooding during Irma.

### Flooding due to storm surge

To estimate storm surge flooding for each storm event we use a 2D depth-averaged hydrodynamic model (hereafter ''flood model''): the Danish Hydraulic Institute (DHI) Mike21 model,[59] a finite volume hydrodynamic model which solves the 2D shallow water equations on an unstructured grid. The model setup consists of a coarse (8–12-km element size), large-scale mesh encompassing the entire Northwest Atlantic Ocean and the Gulf of Mexico, extending landwards up to the US coastline and forced by tides at the open sea boundaries, and a higher-resolution nested mesh with a maximum resolution of 100–200 m at the coastline, extending from the continental shelf break up to inland areas well beyond the furthest modeled flood extent. The nested model is forced at the boundaries using currents and surface elevation extracted from the large-scale model and is run using

Mike21's wetting and drying algorithm enabled. This flood model has previously been presented and validated for Hurricane Sandy in Narayan et al.[14]

Tidal boundary forcing is from the Technical University of Denmark's DTU10 Global Ocean Tidal model, as implemented into the DHI Mike software package.[60] With these boundary conditions, the flood model is used to simulate the propagation of tides and storm surge from the continental shelf onto land during individual storm events. For stochastic storm events, the time of storm genesis is randomly allocated over a 4-week period to ensure sampling of the full tidal cycle. The bathymetry for the model comes from Jeppesen Marine's C-MAP Professional+ digital nautical charts, a global navigational-quality vector chart database,[61] extracted using DHI MIKE software, and land elevation is from the USGS National Elevation Dataset. Land-cover data are obtained from the USGS National Land Cover Database (USGS NLCD), a dataset categorizing land use at 30-m resolution based on remote sensing,[62] and additional data on mangrove extent are obtained from the Florida Fish and Wildlife Commission,[63] the most refined and up-to-date assessment at the time this work was carried out (Figure S4). Mangroves are integrated into the land-cover map under the category of "woody wetlands." These datasets are used to create a spatially varying overland surface roughness field using the Manning roughness coefficient formulation, which is used as an input to the flood model, following USGS guidance.[64]

### Flood model validation

The flood model is used to calculate peak flood depths for Hurricanes Irma and Ian and for the 100 storm events in Collier County. The flood model predicts observed flood depths well during both Hurricanes Irma and Ian with mean errors in modeled high-water marks below 10 cm for both storms and root mean square errors (RMSEs) below 50 cm (Figure 7). For Hurricane Irma, we compare modeled peak surge heights to observed water levels across all US regions, where it caused a storm surge, from Florida to Georgia and parts of South Carolina.

### Property losses from storm surge flooding

For each flood footprint, we estimate economic property losses due to storm surge by integrating spatial data on peak flood depths with property value at that location using depth-damage functions. For Hurricanes Irma and Ian, we estimate property losses due to flooding from these two events for the two mangrove scenarios. In Collier County, we estimate property losses due to the flooding from each of the 100 representative storms and integrate these with information on storm frequencies to obtain an AAL from hurricane-driven coastal flooding for each location. These property losses are estimated for both mangrove scenarios, then we run the models twice for Hurricanes Irma and Ian, and twice for each of the 100 storms in Collier County.

To estimate property losses due to coastal flooding from an event, we first interpolate peak surge heights from the flood model onto a variable resolution grid with a maximum resolution of 100 m in urban areas with high concentration of properties. It is not possible to resolve coastal protection structures less than ~200 m in scale within our hydrodynamic model, due to the constraints imposed by the model's mesh resolution. Therefore, to account for coastal protection unresolved by the model, grid cells behind coastlines with known coastal protection structures such as levees, or behind other artificially defended coastlines like harbors, ports, and so on, are considered not flooded when the flood depths in these cells are below the estimated protection level for that coastline type. The locations of these coastal protection structures are obtained from the US Army Corps of Engineers (USACE) National Levee Database and the FEMA National Flood Hazard Layer datasets. We then apply depth-damage functions to all flooded properties to estimate the economic loss due to flooding. The flood depths are adjusted to account for the protection offered by known, existing coastal protection structures. We then apply damage functions to all flooded properties to estimate the economic damage to different property types based on the flood depths they experience. Waves are accounted for within the loss model using an empirically calibrated, spatially variable multiplication coefficient obtained using Mike21 Spectral Wave model simulations. This acts to increase modeled flood depths where waves are likely to be present by adding 70% of the significant wave height onto the still water depth at these locations to represent the contribution of depth-limited breaking waves to total water depth during the storm (following Federal Emergency Management Agency [FEMA] guidance[65]). Furthermore, the impact of high-energy wave action on structures is modeled by applying a wave-specific vulnerability function to calculate the mean property damage ratio at these locations.

We use a commercial database of property exposure from Moody's RMS that includes information on the structural characteristics and economic value of all insurable residential and commercial properties within the floodplain, i.e., excluding public infrastructure, at high spatial resolution (down to 100 m). These properties include residential dwellings of different types as well as industrial and commercial properties. For each property type, we use calibrated flood depth-damage functions that describe the possible distribution of damage to a structure based on the flood depth and the structure's characteristics, such as the construction type, its occupancy, its height, the year it was built, and whether it has any additional protective features. The damage functions are derived from observations of flood damage compiled and developed by Moody's RMS, calibrated with proprietary data on historic flood insurance claims and structure types, including both physical damage and time element losses such as business interruption (following the approach described in RMS[52]). The time element losses are a function of the level of damage and the building occupancy type (i.e., additional living expenses for residential properties and business interruption for commercial) with higher damage causing higher time element losses. For non-residential buildings, the loss estimates also account for the time element losses caused by the loss of function of critical lifeline systems such as power networks, water, or waste management systems. At all locations affected by coastal waves, the damage functions account for the presence of high-energy wave action. Thus, we obtain spatially variable property losses from storm surges associated with specific events. All losses reported here are without any insurance terms such as deductibles or limits applied. All losses are estimated in terms of 2018 US$.

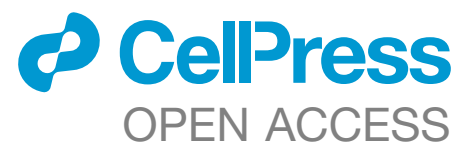

### Storm event set for Collier County

We start with the full US-wide synthetic tropical cyclone (TC) track set described in hurricane parameters, which represents 100,000 years of TC activity over the US Atlantic and Gulf coasts, and then select only the TC events causing any loss in Collier County, which results in 3,966 TC events. Due to the high computational cost of running many thousands of flood simulations for both With-Mangrove and No-Mangrove scenarios, we then reduce this full event set into a smaller set of 100 representative events, which produces the same county-wide AAL as the full set. To obtain these 100 representative storm events for Collier County, we first obtain the full event loss set by calculating location-specific storm surge loss, for every property within the property exposure database, for all 3,966 synthetic storm events impacting the county. Each event in this set has an assigned frequency, or rate, calibrated to the observed frequency of storms in the county for the period AD 1900–2011. The AAL contribution from each event is the product of the loss from the event and its rate. To calculate the overall AAL, we sum the AAL contributions from every event.

We then divide this full event set into 100 equally spaced AAL quantiles, such that the total AAL for each quantile is the sum of the AAL contributions from all events in this quantile. We then select a single event from each quantile, with an assumed rate equal to the sum of all event rates in this quantile. This event is selected such that its loss contribution (i.e., event loss multiplied by new rate) is closest to the total AAL for its quantile.

Thus, the subsampled set is produced to match the distribution of losses from the full set as closely as possible, and the AAL values from the subsampled event set match the AAL values from the full event set closely (see error analysis in Table S1; Figure 1). To calculate the overall AAL, we sum the AAL contributions from all events. AALs are then computed and compared with the With-Mangroves and No-Mangrove scenarios.

Errors in AAL values are defined as the difference between the full event set and subsampled event set. County-wide, the subsampled event set with 100 events was found to very closely approximate the full event set in terms of AAL, with an average AAL error across all storm events of +0.045%, and a standard deviation error of +0.008%, meaning that biases in both the AAL and variance around it are negligible (Table S1). To further evaluate potential biases introduced by the event set reduction for events with an RP < 10,000 years, we calculate a metric we define as the inverse excess AAL (IXSAAL), which is the inverse of the excess AAL (or XSAAL) metric that is commonly used in catastrophe risk modeling and quantifies the AAL for events whose loss RP at or above a given value (see FEMA[65]). The IXSAAL is calculated as AAL−XSAAL and quantifies the cumulative contribution to the total AAL from events whose loss RP is lower than a specified level. We find that AAL in Collier County is mainly driven by events with an RP < 100 years; in terms of total damages caused, events with an RP < 100 years account for 56% of the total AAL, and events with an RP > 10,000 years account for 1% of the total AAL (Table S1). The errors within this range in the subsampled set are low: −0.25% error in IXSAAL for events with an RP < 10,000 years and +0.071% error for events with an RP < 100 years.

For the mangrove benefit analyses, we aggregate AAL values by summing values within 10-km$^2$ hexagons. We examine errors introduced by this aggregation by comparing total AAL values by hexagon across the full and subsampled event set ignoring hexagons where total AAL values from both full set and subsampled set are < US$100. The AAL values from our subsampled set are very close to values from the full set ($R^2$ = 0.999) (Figure S1).

### Effect of mangroves on storm surge flooding and property loss

The effects of mangroves are measured in terms of the spatial differences in peak flood extents and depths, and subsequent differences in property losses from storm surge flooding for each storm event between two scenarios: "With Mangroves" and "No Mangroves" within the flood model. In the With-Mangroves scenario, mangrove effects on flooding are represented through a Manning's friction coefficient of 0.1. In the No-Mangroves scenario, all mangroves within the model domain are re-classified with a reduced Manning's friction coefficient of 0.02, corresponding to a smoother surface characteristic of the friction produced by developed open space, or open-water seabeds,[21] while all other land-cover and model conditions remain unchanged. This approach allows us to isolate the influence of the presence of mangrove vegetation on flood extents. The choice of coefficient for the With-Mangroves scenario follows previous studies[23] and results in good validation with observed overland high-water marks (see Figure 7). This value is lower than the most commonly used values, 0.14–0.15, within 2D modeling studies.[24,26] Thus, the modeled flood depths in this study can be interpreted as a conservative estimate of the effects of mangroves.

The effects of mangroves on storm surge property damages are then calculated as the difference in property damages between the flood extents in the With-Mangroves and No-Mangroves scenarios. For Hurricanes Irma and Ian, mangrove effects are calculated as the difference between storm surge damages from each hurricane event for the two mangrove scenarios. In the stochastic event study for Collier County, we estimate the AAL values for both mangrove scenarios for 100 storm events, calculate the difference in AAL values between the two scenarios for each event, and sum these difference values to obtain the total AAL reduction benefits of mangroves.

To understand the types of storm events against which mangroves provide the most flood reduction benefits, we assess the proportion of total mangrove benefits received by return period (RP). To do this, we plot the cumulative distribution function (CDF) of mangrove AAL benefits normalized by total mangrove AAL benefits, against event RP. All spatial analyses were conducted in ArcGIS and all data analyses in Excel and R.

We characterize the spatial patterns of mangrove effects on storm surge damages to properties in southern Florida for all three cases: AALs, Ian, and Irma. We first manually define the landward mangrove edge, i.e., the edge demarcating the landward-most extent of mangroves in ArcGIS. We then aggregate all property loss values into hexagonal units, each unit having a side of 10 km. Next, we classify each hexagon as either landward of the mangrove edge, if it lies completely landward of the mangrove edge, or between/inside the mangroves if it intersects or lies seaward of this edge. We then calculate the shortest distance of the center of each hexagon to the landward mangrove edge in kilometers. For Irma, we exclude properties in central Florida and the lower Florida Keys in these analyses due to the small number of mangroves and the small difference in flood depths due to the mangroves in these regions (Figure S2).

Finally, we calculate the range, mean, median, and quartile values of positive and negative mangrove effects on property damages for all hexagonal units within 1-km bins from the forest edge until the most distant hexagon. For each of the three cases, i.e., AALs, Irma, or Ian, we consider mangrove effects to be significant in a hexagon if the maximum value of mangrove effects in that hexagon is greater than 1% of the maximum value across all hexagons in that case. All economic values are in 2018 US$. Further details regarding the methods can be found in the supplemental information.

## RESOURCE AVAILABILITY

### Lead contact

Requests for further information and resources should be directed to and will be fulfilled by the lead contact, Siddharth Narayan (narayans19@ecu.edu).

### Materials availability

This study did not generate new unique reagents.

### Data and code availability

- All model generated data supporting the results and figures presented in this study will be shared by the lead contact upon request. These data will include (1) CSV or shapefile of average flood depths by 10-km$^2$ hexagon, for each scenario for Hurricanes Irma and Ian; (2) CSV of % difference in flood damages between No-Mangroves and With-Mangroves scenarios by hexagon, for Irma and for the Collier County Annual Average Loss analyses; and (3) CSV of absolute change in XSAAL and inverse XSAAL at event RPs from 15 to 1,000 years. Certain underlying datasets, such as property-level loss values and calibrated damage functions, are proprietary data that are central to the business interests of Moody's RMS, and as such cannot be shared publicly. All other data pertaining to mangrove extents and county boundaries are publicly available and appropriately cited in the manuscript.
- This study reports no original code.
- Any additional information required to reanalyze the data reported in this paper is available from the lead contact upon request.

## ACKNOWLEDGMENTS

This work was supported by funding from the Walton Family Foundation and the Herbert. W. Hoover Foundation. S.N. was also supported by the National Science Foundation through NSF award # 2206479. M.W.B. was also supported by the AXA Research Fund and NSF award # 2209284. The graphical abstract for this paper was created by Jessica Kendall-Bar. We gratefully acknowledge the thorough comments from our anonymous reviewers and the efforts of the editor during this process.

## AUTHOR CONTRIBUTIONS

Conceptualization, S.N., C.J.T., and M.W.B.; methodology, S.N. and C.J.T.; investigation, S.N. and C.J.T.; writing—original draft, all authors; writing—review and editing, all authors; funding acquisition, S.N., C.S., and M.W.B.

## DECLARATION OF INTERESTS

C.J.T. and K.N. are employees of Moody's RMS, London. C.S. and L.G. are employees of The Nature Conservancy, USA. J.M. was an employee at Moody's RMS and is currently an employee at Reask, London.

## SUPPLEMENTAL INFORMATION

Supplemental information can be found online at https://doi.org/10.1016/j.crsus.2025.100531

## REFERENCES

1. Maza, M., Lara, J.L., and Losada, I.J. (2019). Experimental analysis of wave attenuation and drag forces in a realistic fringe Rhizophora mangrove forest. Adv. Water Resour. *131*, 103376. https://doi.org/10.1016/j.advwatres.2019.07.006.
2. Sánchez-Núñez, D.A., Bernal, G., and Pineda, J.E.M. (2019). The relative role of mangroves on wave erosion mitigation and sediment properties. Estuaries Coasts *42*, 2124–2138. https://doi.org/10.1007/s12237-019-00628-9.
3. Montgomery, J.M., Bryan, K.R., Mullarney, J.C., and Horstman, E.M. (2019). Attenuation of storm surges by coastal mangroves. Geophys. Res. Lett. *46*, 2680–2689. https://doi.org/10.1029/2018GL081636.
4. Menéndez, P., Losada, I.J., Torres-Ortega, S., Narayan, S., and Beck, M. W. (2020). The global flood protection benefits of mangroves. Sci. Rep. *10*, 4404. https://doi.org/10.1038/s41598-020-61136-6.
5. Hochard, J.P., Hamilton, S., and Barbier, E.B. (2019). Mangroves shelter coastal economic activity from cyclones. Proc. Natl. Acad. Sci. USA *116*, 12232–12237. https://doi.org/10.1073/pnas.1820067116.
6. Sun, F., and Carson, R.T. (2020). Coastal wetlands reduce property damage during tropical cyclones. Proc. Natl. Acad. Sci. USA *117*, 5719–5725. https://doi.org/10.1073/pnas.1915169117.
7. Menéndez, P., Losada, I.J., Beck, M.W., Torres-Ortega, S., Espejo, A., Narayan, S., Díaz-Simal, P., and Lange, G.M. (2018). Valuing the protection services of mangroves at national scale: The Philippines. Ecosyst. Serv. *34*, 24–36. https://doi.org/10.1016/j.ecoser.2018.09.005.
8. Bucci, L., Alaka, L., Hagen, A., Delgado, S., and Beven, J. (2023). National Hurricane Center Tropical Cyclone Report. Hurricane Ian (AL092022), pp. 1–72. https://www.nhc.noaa.gov/data/tcr/AL092022_Ian.pdf.
9. Blake, E.S. (2018). The 2017 Atlantic hurricane season: catastrophic losses and costs. Weatherwise *71*, 28–37. https://doi.org/10.1080/00431672.2018.1448147.
10. Cangialosi, J.P., Latto, A.S., and Berg, R. (2021). National Hurricane Center Tropical Cyclone Report. Hurricane Irma (AL112017). https://www.nhc.noaa.gov/data/tcr/AL112017_Irma.pdf.
11. Costanza, R., Anderson, S.J., Sutton, P., Mulder, K., Mulder, O., Kubiszewski, I., Wang, X., Liu, X., Pérez-Maqueo, O., Luisa Martinez, M., et al. (2021). The global value of coastal wetlands for storm protection. Glob. Environ. Change *70*, 102328. https://doi.org/10.1016/j.gloenvcha.2021.102328.
12. Al-Attabi, Z., Xu, Y., Tso, G., and Narayan, S. (2023). The impacts of tidal wetland loss and coastal development on storm surge damages to people and property: A Hurricane Ike case-study. Sci. Rep. *13*, 4620. https://doi.org/10.1038/s41598-023-31409-x.
13. Taylor-Burns, R., Lowrie, C., Tehranirad, B., Lowe, J., Erikson, L., Barnard, P.L., Reguero, B.G., and Beck, M.W. (2024). The value of marsh restoration for flood risk reduction in an urban estuary. Sci. Rep. *14*, 6856. https://doi.org/10.1038/s41598-024-57474-4.

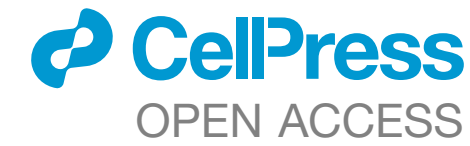

14. Narayan, S., Beck, M.W., Wilson, P., Thomas, C.J., Guerrero, A., Shepard, C.C., Reguero, B.G., Franco, G., Ingram, J.C., and Trespalacios, D. (2017). The value of coastal wetlands for flood damage reduction in the northeastern USA. Sci. Rep. *7*, 9463. https://doi.org/10.1038/s41598-017-09269-z.
15. US Coral Reef Task Force (2023). USCRTF Resolution 47.2: Coral Reefs as National Natural Infrastructure. https://taskforce.coralreef.noaa.gov/assets/meeting47/pdf/Resolution47.2_Coral-Reefs-As-Natural-Coastal-Infrastructure_10-20-23.pdf.
16. Pinsky, M.L., Guannel, G., and Arkema, K.K. (2013). Quantifying wave attenuation to inform coastal habitat conservation. Ecosphere *4*, 1–16. https://doi.org/10.1890/ES13-00080.1.
17. Loder, N.M., Irish, J.L., Cialone, M.A., and Wamsley, T.V. (2009). Sensitivity of hurricane surge to morphological parameters of coastal wetlands. Estuarine Coastal Shelf Sci. *84*, 625–636. https://doi.org/10.1016/j.ecss.2009.07.036.
18. Gittman, R.K., Scyphers, S.B., Baillie, C.J., Brodmerkel, A., Grabowski, J. H., Livernois, M., Poray, A.K., Smith, C.S., and Fodrie, F.J. (2021). Reversing a tyranny of cascading shoreline-protection decisions driving coastal habitat loss. Conserv. Sci. Pract. *3*, e490. https://doi.org/10.1111/csp2.490.
19. Guthrie, A.G., Stafford, S., Scheld, A.M., Nunez, K., and Bilkovic, D.M. (2023). Property owner shoreline modification decisions vary based on their perceptions of shoreline change and interests in ecological benefits. Front. Mar. Sci. *10*, 1031012. https://doi.org/10.3389/fmars.2023.1031012.
20. Kok, S., Bisaro, A., de Bel, M., Hinkel, J., and Bouwer, L.M. (2021). The potential of nature-based flood defences to leverage public investment in coastal adaptation: Cases from the Netherlands, Indonesia and Georgia. Ecol. Econ. *179*, 106828. https://doi.org/10.1016/j.ecolecon.2020.106828.
21. Liu, H., Zhang, K., Li, Y., and Xie, L. (2013). Numerical study of the sensitivity of mangroves in reducing storm surge and flooding to hurricane characteristics in southern Florida. Contin. Shelf Res. *64*, 51–65. https://doi.org/10.1016/j.csr.2013.05.015.
22. Deb, M., and Ferreira, C.M. (2017). Potential impacts of the Sunderban mangrove degradation on future coastal flooding in Bangladesh. J. Hydro-Environ. Res. *17*, 30–46. https://doi.org/10.1016/j.jher.2016.11.005.
23. Mattocks, C., and Forbes, C. (2008). A real-time, event-triggered storm surge forecasting system for the state of North Carolina. Ocean Modell. *25*, 95–119. https://doi.org/10.1016/j.ocemod.2008.06.008.
24. Xu, H., Zhang, K., Shen, J., and Li, Y. (2010). Storm surge simulation along the U.S. East and Gulf Coasts using a multi-scale numerical model approach. Ocean Dyn. *60*, 1597–1619. https://doi.org/10.1007/s10236-010-0321-3.
25. Bryant, M.A., Bryant, D.B., Provost, L.A., Hurst, N., McHugh, M., Wargula, N., and Tomiczeck, T. (2022). Wave attenuation of coastal mangroves at a near-prototype scale. U.S. Army Corps of Engineers (ERDC). https://apps.dtic.mil/sti/pdfs/AD1181016.pdf.
26. Zhang, K., Liu, H., Li, Y., Xu, H., Shen, J., Rhome, J., and Smith, T.J. (2012). The role of mangroves in attenuating storm surges. Estuarine Coastal Shelf Sci. *102*, 11–23. https://doi.org/10.1016/j.ecss.2012.02.021.
27. Chen, Q., Li, Y., Kelly, D.M., Zhang, K., Zachry, B., and Rhome, J. (2021). Improved modeling of the role of mangroves in storm surge attenuation. Estuarine Coastal Shelf Sci. *260*, 107515. https://doi.org/10.1016/j.ecss.2021.107515.
28. Sheng, Y.P., and Zou, R. (2017). Assessing the role of mangrove forest in reducing coastal inundation during major hurricanes. Hydrobiologia *803*, 87–103. https://doi.org/10.1007/s10750-017-3201-8.
29. National Centers for Coastal Ocean Science. (2025). RESTORE Sponsored Research Project: A Web-Based Interactive Decision-Support Tool for Adaptation of Coastal Urban and Natural Ecosystems (ACUNE) in Southwest Florida. https://www.fisheries.noaa.gov/inport/item/58981.
30. Sheng, Y.P., Rivera-Nieves, A.A., Zou, R., and Paramygin, V.A. (2021). Role of wetlands in reducing structural loss is highly dependent on characteristics of storms and local wetland and structure conditions. Sci. Rep. *11*, 5237. https://doi.org/10.1038/s41598-021-84701-z.
31. Tomiczek, T., Wargula, A., Lomonaco, P., Goodwin, S., Cox, D., Kennedy, A., and Lynett, P. (2019). Physical model investigation of parcel-scale effects of mangroves on wave transformation and force reduction in the built environment. Coastal Eng. *2020*, 998–1007. https://doi.org/10.9753/icce.v36v.waves.1.
32. Thomas, N., Lucas, R., Bunting, P., Hardy, A., Rosenqvist, A., and Simard, M. (2017). Distribution and drivers of global mangrove forest change, 1996–2010. PLoS one *12*, e0179302. https://doi.org/10.1371/journal.pone.0179302.
33. Association of State Floodplain Managers. (2022). Estimated damage for Hurricane Ian Between 41 Billion and 70 Billion. https://www.floods.org/news-views/research-and-reports/estimated-damage-for-hurricane-ian-between-41-billion-and-70-billion/.
34. NCEI. (2022). Storm Events Database. https://www.ncdc.noaa.gov/stormevents/.
35. NCEI. (2022). U.S. Billion-Dollar Weather and Climate Disasters. https://www.ncei.noaa.gov/access/billions/.
36. FEMA. (2019). Hurricane Irma Flood Insurance Payments Top $1 Billion in Florida. https://www.fema.gov/press-release/20210318/hurricane-irma-flood-insurance-payments-top-1-billion-florida.
37. Michel-Kerjan, E.O., and Kousky, C. (2010). Come rain or shine: Evidence on flood insurance purchases in Florida. J. Risk Ins. *77*, 369–397. https://doi.org/10.1111/j.1539-6975.2009.01349.x.
38. Tomiczek, T., O'Donnell, K., Furman, K., Webbmartin, B., and Scyphers, S. (2020). Rapid Damage Assessments of Shorelines and Structures in the Florida Keys after Hurricane Irma. Nat. Hazards Rev. *21*, 05019006. https://doi.org/10.1061/(ASCE)NH.1527-6996.0000349.
39. Montgomery, M., and Kunreuther, H. (2018). Pricing storm surge risks in Florida: Implications for determining flood insurance premiums and evaluating mitigation measures. Risk Anal. *38*, 2275–2299. https://doi.org/10.1111/risa.13127.
40. Lagomasino, D., Fatoyinbo, T., Castañeda-Moya, E., Cook, B.D., Montesano, P.M., Neigh, C.S.R., Corp, L.A., Ott, L.E., Chavez, S., and Morton, D. C. (2021). Storm surge and ponding explain mangrove dieback in southwest Florida following Hurricane Irma. Nat. Commun. *12*, 4003. https://doi.org/10.1038/s41467-021-24253-y.
41. Temmerman, S., Horstman, E.M., Krauss, K.W., Mullarney, J.C., Pelckmans, I., and Schoutens, K. (2023). Marshes and mangroves as nature-based coastal storm buffers. Annu. Rev. Mar. Sci. *15*, 95–118. https://doi.org/10.1146/annurev-marine-040422-092951.
42. Lange, G.-M., Beck, M.W., Lam, V., Menendez, P., and Sumaila, R. (2021). Blue Natural Capital: Mangroves and Fisheries (World Bank), pp. 121–143. https://doi.org/10.1596/978-1-4648-1590-4_ch6.
43. Brody, S.D., Zahran, S., Maghelal, P., Grover, H., and Highfield, W.E. (2007). The Rising Costs of Floods: Examining the Impact of Planning and Development Decisions on Property Damage in Florida. J. Am. Plann. Assoc. *73*, 330–345. https://doi.org/10.1080/01944360708977981.
44. Beck, M.W., Heck, N., Narayan, S., Menéndez, P., Reguero, B.G., Bitterwolf, S., Torres-Ortega, S., Lange, G.M., Pfliegner, K., Pietsch McNulty, V., et al. (2022). Return on Investment for Mangrove and Reef Flood Protection. Ecosyst. Serv. *56*, 101440. https://doi.org/10.1016/j.ecoser.2022.101440.
45. Narayan, S., Thomas, C., Matthnewman, J., Shepard, C.C., Geselbracht, L., Nzerem, K., and Beck, M.W. (2019). Valuing the Flood Risk Reduction Benefits of Florida's Mangroves. Conservation Gateway. https://www.nature.org/content/dam/tnc/nature/en/documents/Mangrove_Report_digital_FINAL.pdf.
46. National Hurricane Center and Central Pacific Hurricane Center. (2024). Storm Surge Overview. https://www.nhc.noaa.gov/surge/.
47. Airoldi, L., Beck, M.W., Firth, L.B., Bugnot, A.B., Steinberg, P.D., and Dafforn, K.A. (2021). Emerging solutions to return nature to the urban ocean.

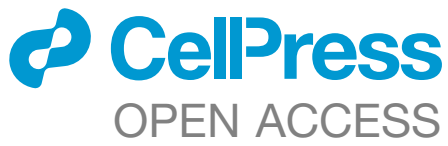

Annu. Rev. Mar. Sci. *13*, 445–477. https://doi.org/10.1146/annurev-marine-032020-020015.

48. Narayan, S., Nicholls, R.J., Clarke, D., Hanson, S., Reeve, D., Horrillo-Caraballo, J., le Cozannet, G., Hissel, F., Kowalska, B., Parda, R., et al. (2014). The SPR systems model as a conceptual foundation for rapid integrated risk appraisals: Lessons from Europe. Coastal Eng. *87*, 15–31. https://doi.org/10.1016/j.coastaleng.2013.10.021.

49. Jarvinen, B.R., Neumann, C.J., and Davis, M.A. (1984). A tropical cyclone data tape for the North Atlantic Basin, 1886-1983: Contents, limitations, and uses. https://repository.library.noaa.gov/view/noaa/7069.

50. Hall, T.M., and Jewson, S. (2022). Statistical modelling of North Atlantic tropical cyclone tracks. Tellus A *59*, 486–498. https://doi.org/10.1111/j.1600-0870.2007.00240.x.

51. Hall, T.M., and Jewson, S. (2008). Comparison of local and basinwide methods for risk assessment of tropical cyclone landfall. J. Appl. Meteorol. Climatol. *47*, 361–367. https://doi.org/10.1175/2007JAMC1720.1.

52. RMS (2019). Submission to the Florida Commission on Hurricane Loss Projection Methodology: North Atlantic Hurricane Models Version 18.1 (Build 1945) (Risk Management Solutions). https://fchlpm.sbafla.com/media/dnbbgta1/rms17standardsanddisclosures_05212019.pdf.

53. Willoughby, H.E., Darling, R.W.R., and Rahn, M.E. (2006). Parametric representation of the primary hurricane vortex. Part II: A new family of sectionally continuous profiles. Mon. Weather Rev. *134*, 1102–1120. https://doi.org/10.1175/MWR3106.1.

54. Demuth, J.L., DeMaria, M., and Knaff, J.A. (2006). Improvement of Advanced Microwave Sounding Unit tropical cyclone intensity and size estimation algorithms. J. Appl. Meteorol. Climatol. *45*, 1573–1581. https://doi.org/10.1175/JAM2429.1.

55. Powell, M.D., Houston, S.H., and Reinhold, T.A. (1996). Hurricane Andrew's landfall in south Florida. Part I: Standardizing measurements for documentation of surface wind fields. Weather Forecasting *11*, 304–328. https://doi.org/10.1175/1520-0434(1996)011<0304:HALISF>2.0.CO;2.

56. Powell, M.D., Houston, S.H., Amat, L.R., and Morisseau-Leroy, N. (1998). The HRD real-time hurricane wind analysis system. J. Wind Eng. Ind. Aerodyn. *77–78*, 53–64. https://doi.org/10.1016/S0167-6105(98)00131-7.

57. Powell, M.D., Murillo, S., Dodge, P., Uhlhorn, E., Gamache, J., Cardone, V., Cox, A., Otero, S., Carrasco, N., Annane, B., et al. (2010). Reconstruction of Hurricane Katrina's wind fields for storm surge and wave hindcasting. Ocean Eng. *37*, 26–36. https://doi.org/10.1016/j.oceaneng.2009.08.014.

58. Holland, G.J. (1980). An analytic model of the wind and pressure profiles in hurricanes. Mon. Wea. Rev. *108*, 1212–1218. https://doi.org/10.1175/1520-0493(1980)108<1212:AAMOTW>2.0.CO;2.

59. Warren, I.R., and Bach, H.K. (1992). MIKE 21: a modelling system for estuaries, coastal waters and seas. Environ. Software *7*, 229–240. https://doi.org/10.1016/0266-9838(92)90006-P.

60. Danish; Hydraulic Institute (2017). Global Tide Model–Tidal Prediction (Danish Hydraulic Institute).

61. Danish Hydraulic Institute. (2021). C-MAP. https://www.c-map.com/home/.

62. Homer, C., Dewitz, J., Yang, L., Jin, S., Danielson, P., Xian, G., Coulston, J., Herold, N., Wickham, J., and Megown, K. (2015). Completion of the 2011 National Land Cover Database for the conterminous United States–representing a decade of land cover change information. Photogramm. Eng. Remote Sens. *81*, 345–354. https://doi.org/10.1016/S0099-1112(15)30100-2.

63. FWC. (2019). Mangrove Habitat in Florida. https://geodata.myfwc.com/datasets/myfwc::mangrove-habitat-in-florida-1/about.

64. Arcement, G.J., and Schneider, V.R. (1989). Guide for Selecting Manning's Roughness Coefficients for Natural Channels and Flood Plains (United States Geological Survey Water-Supply Paper 2339). https://pubs.usgs.gov/wsp/2339/report.pdf.

65. FEMA (2011). FEMA P-55: Coastal Construction Manual—Principles and Practices of Planning, Siting, Designing, Constructing, and Maintaining Residential Buildings in Coastal Areas (Federal Emergency Management Agency). https://www.fema.gov/sites/default/files/2020-08/fema55_volii_combined_rev.pdf.

## Supplemental information

# The spatially variable effects of mangroves on flood depths and losses from storm surges in Florida

**Siddharth Narayan, Christopher J. Thomas, Kechi Nzerem, Joss Matthewman, Christine Shepard, Laura Geselbracht, and Michael W. Beck**

**Supplementary Information: The spatially variable effects of mangroves on flood depths and losses from storm surges in Florida**

1. AAL Storm Event Set Biases and Errors

Table S1: **Comparison of AAL values of the limited event set used in this study with the full storm event set for Collier County**

| Return Period (yrs) | InvXSAAL Error (this study - full set) | Relative Event Contribution to AAL (i.e. $InvXSAAL_{FullSet}$ / $AAL_{FullSet}$ ) |
|---|---|---|
| 50 | -0.071% | 34% |
| 100 | +0.071% | 56% |
| 1K | +0.057% | 92% |
| 10K | -0.247% | 99% |
| 100K | +0.122% | >99.999% |

For the mangrove benefit analyses we aggregate AAL values by summing values within 10 $km^2$ hexagonal teselas. We examine errors introduced by this aggregation by comparing total AAL values by tesela across the full and subsampled event set ignoring teselas where total AAL values from both full set and subsampled set are < $100. The AAL values from our subsampled set are very close to values from the full set ($R^2$ = 0.999). (Supplementary Figure 1).

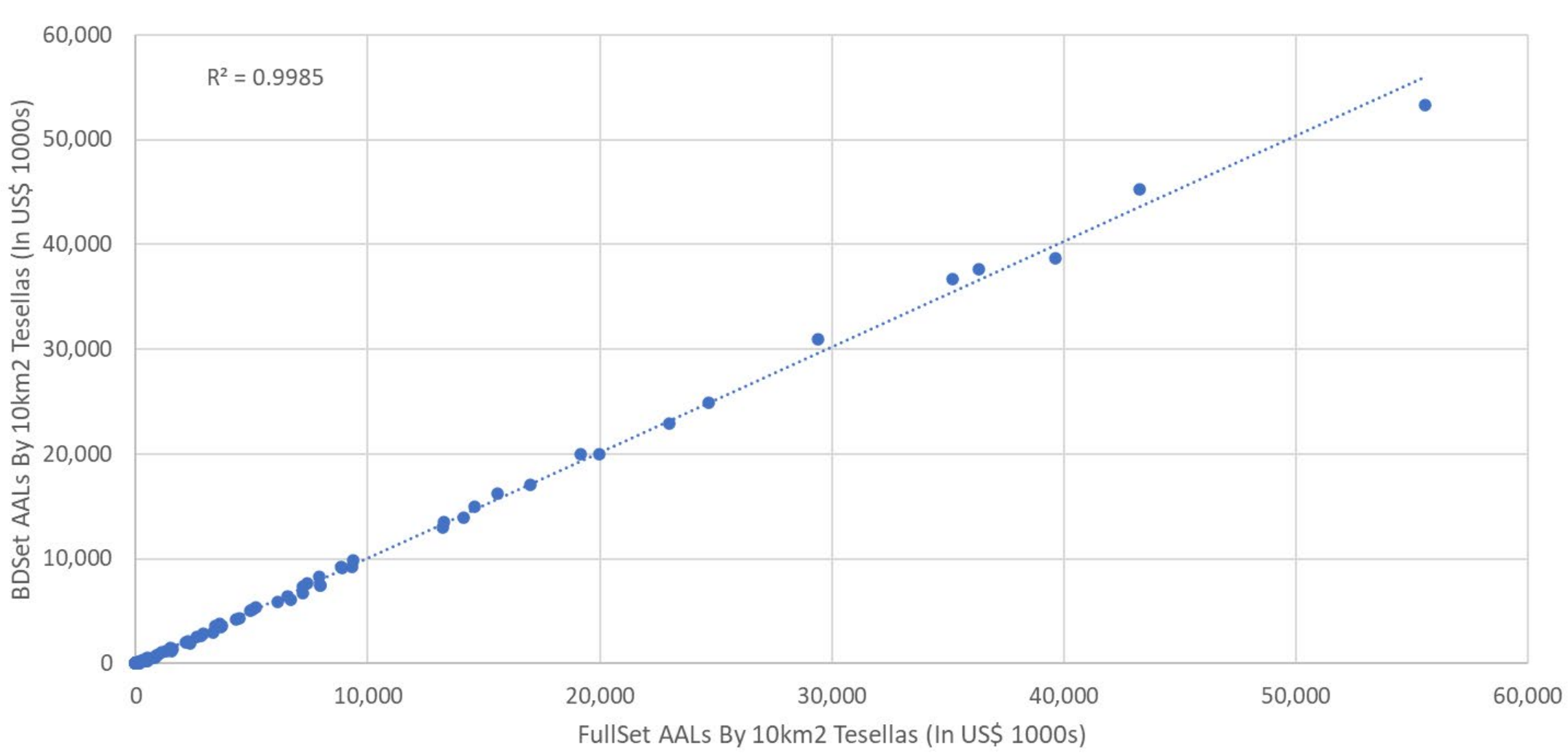

Figure S1: **Scatter-plot of total AAL values by tesela for fullset AALs versus subsampled set AALs**

## SI 2. Flood depths and damages

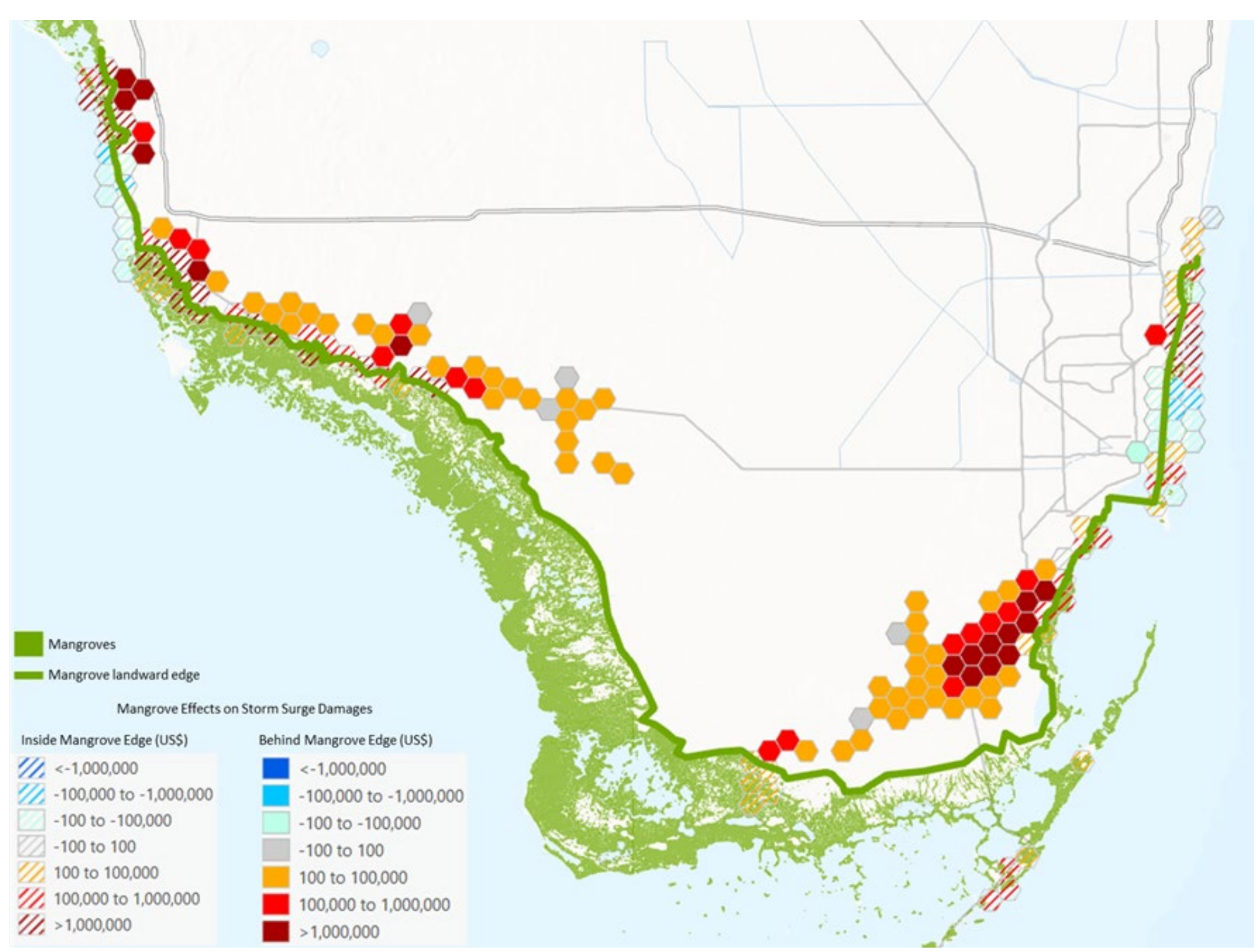

Figure S2: **Difference in storm surge damages between the No Mangrove and With Mangrove scenarios for Hurricane Irma.** Mangroves are in green and Landward mangrove edge is shown as a dark green line.

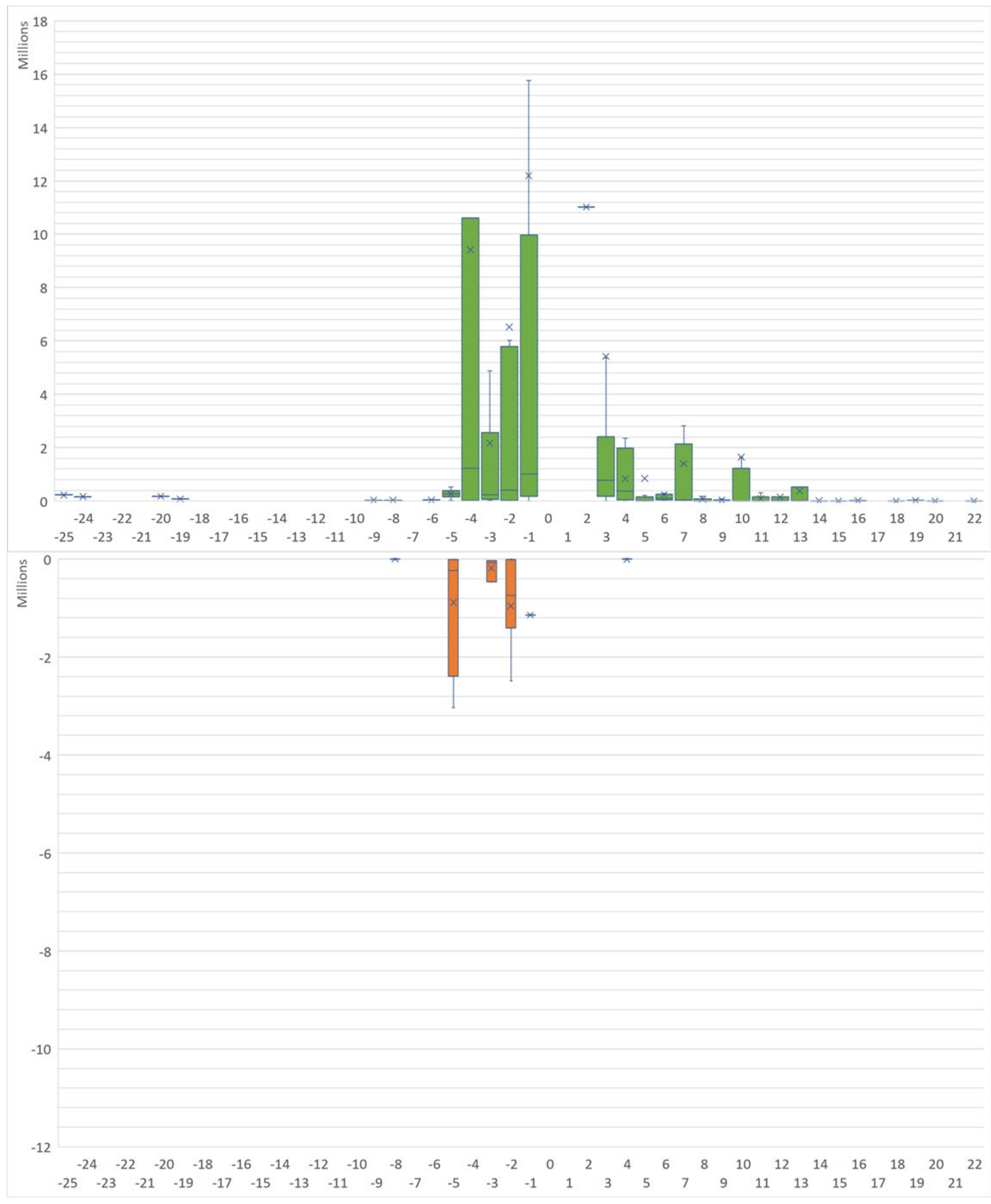

Figure S3: **Box and whisker plot showing difference in storm surge damages (Million US$) between With Mangrove and No Mangrove**. Graph shows 1-km distance bins from the landward edge of the mangrove forests for: A (Top): Collier County AALs and B (Bottom): Hurricane Irma. Orange bars show reduction in risk due to mangroves and blue bars show increase in risk due to mangroves. Green line indicates landward mangrove edge.

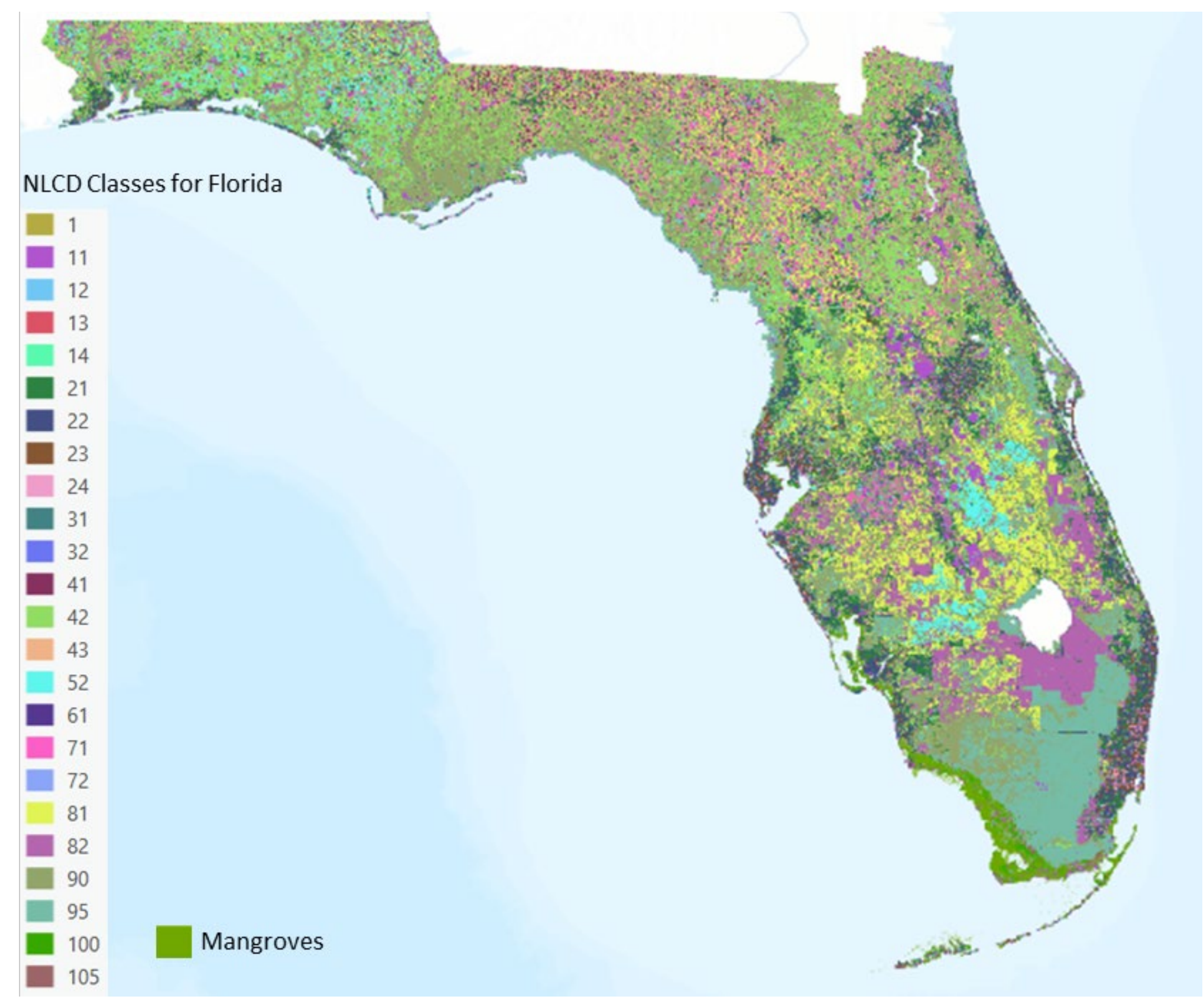

Figure S4: **Map of land cover classes in Florida from the National Land Cover Database (NLCD).** Manning's coefficients for these land-covers are based on Zhang et al. (2012) as described in Methods in the Main article.